\documentclass{aims} 
\usepackage{amsmath}
\usepackage{paralist}
\usepackage[misc]{ifsym}
\usepackage{epsfig} 
\usepackage{epstopdf} 
\usepackage[colorlinks=true]{hyperref}
\hypersetup{urlcolor=blue, citecolor=red}
\allowdisplaybreaks

\def\currentvolume{X}
 \def\currentissue{X}
  \def\currentyear{200X}
   \def\currentmonth{XX}
    \def\ppages{X-XX}
     \def\DOI{10.3934/xx.xxxxxxx}

\usepackage{array,multirow}
\usepackage[table,xcdraw]{xcolor}
\usepackage{tabularx}
\usepackage{subfigure}
\usepackage{amssymb}
\usepackage{enumitem}
\usepackage{grffile}
\usepackage{tabularx}
\usepackage{soul}
\usepackage{booktabs}
\usepackage[figuresright]{rotating}
\usepackage[pagewise]{lineno}

\newtheorem{theorem}{Theorem}[section]

\theoremstyle{definition}
\newtheorem{definition}[theorem]{Definition}

\usepackage{floatrow}
\newfloatcommand{capbtabbox}{table}[][\FBwidth]
\newcolumntype{H}{>{\setbox0=\hbox\bgroup}c<{\egroup}@{}}

\usepackage[abs]{overpic}  

\newcommand{\overimgx}[3][]{%
	\begin{overpic}[#1]{#2}%
		\put (2, 2) {%
			\setlength{\fboxsep}{2pt}%
			\colorbox{white}{%
				\scriptsize\sffamily\vphantom{y}%
				#3%
			}%
		}%
	\end{overpic}%
}

\newcommand{\overimgxx}[3][]{%
	\begin{overpic}[#1]{#2}%
		\put (50, 2) {%
			\setlength{\fboxsep}{2pt}%
			\colorbox{white}{%
				\scriptsize\sffamily\vphantom{y}%
				#3%
			}%
		}%
	\end{overpic}%
}

\usepackage{highlight}
\definecolor{high}{HTML}{ef3b2c} 
\definecolor{mid}{HTML}{fff7f7}  
\definecolor{low}{HTML}{6871ff}  
\newcommand{\g}[1]{\gradientcelld{#1}{0}{150}{980}{low}{mid}{high}{90}}
\newcommand{\gl}[1]{\gradientcelld{#1}{0}{0.5}{1}{low}{mid}{high}{90}}

\newcommand{\A}{DN}
\newcommand{\B}{DM}

\title[How to Best Combine Demosaicing and Denoising?]
{How to Best Combine Demosaicing and Denoising?} 

\author[Yu Guo, Qiyu Jin,Jean-Michel Morel and Gabriele Facciolo]{}


\thanks{$^*$Corresponding author: Qiyu Jin}

\begin{document}
\maketitle

    \centerline{\scshape
    Yu Guo$^{{\href{mailto:yuguomath@aliyun.com}{\textrm{\Letter}}}1}$
    , Qiyu Jin$^{{\href{mailto:qyjin2015@aliyun.com}{\textrm{\Letter}}}*1}$
    , Jean-Michel Morel$^{{\href{mailto:jeamorel@cityu.edu.hk}{\textrm{\Letter}}}2}$
    and Gabriele Facciolo$^{{\href{mailto:gabriele.facciolo@ens-paris-saclay.fr}{\textrm{\Letter}}}3}$ }

\medskip

{\footnotesize
 \centerline{$^1$School of Mathematical Science, Inner Mongolia University, Hohhot 010020, China}
} 

\medskip

{\footnotesize
 \centerline{$^2$Department of Mathematics, City University of Hong Kong, Kowloon Tong, Hong Kong}
} 

\medskip

{\footnotesize
 \centerline{$^3$Centre Borelli, ENS Paris-Saclay, CNRS, 4, avenue des Sciences 91190 Gif-sur-Yvette, France}
}

\bigskip

 \centerline{(Communicated by Handling Editor)}


\begin{abstract}
    Image demosaicing and denoising play a critical role in the raw imaging pipeline. These processes have often been treated as independent, without considering their interactions. Indeed, most classic denoising methods handle noisy RGB images, not raw images. Conversely, most  demosaicing methods address the demosaicing of noise free images. The real problem is to jointly denoise and demosaic noisy raw images. But the question of how to proceed is still not yet clarified. In this paper, we carry-out extensive experiments and a mathematical analysis to tackle this problem by low complexity algorithms. Indeed, both problems have been only addressed jointly by end-to-end heavy weight convolutional neural networks (CNNs), which are currently incompatible with low power portable imaging devices and remain by nature domain (or device) dependent. Our study leads us to conclude that, with moderate noise, demosaicing should be applied first, followed by  denoising. This requires a simple adaptation of classic denoising algorithms to  demosaiced noise, which we justify and specify. Although our main conclusion is ``demosaic first, then denoise'', we also discover that for high noise, there is a moderate PSNR gain by a more complex strategy: partial CFA denoising followed by demosaicing, and by a second denoising on the RGB image. These surprising results are obtained by a black-box optimization of the pipeline, which could  be applied to any other pipeline. We validate our results on simulated and real noisy CFA images obtained from several benchmarks.
\end{abstract}


\section{Introduction}
Most portable digital imaging devices acquire images as mosaics, with a color filter array (CFA), sampling only one color value for each pixel. 
The most popular CFA is the Bayer color array~\cite{bayer1976color} where two out of four pixels measure the green (G) value, one measures the red (R) and one the blue (B). 
The two missing color values at each pixel need to be estimated for reconstructing a complete image from a CFA image. 
The process is commonly referred to as CFA interpolation or demosaicing. CFA images have noise, especially in low light conditions, so denoising is also a key step in the imaging pipeline.

Denoising and demosaicing are often handled as two independent operations~\cite{paliy2007demosaicing}  for processing noisy raw sensor data.
Most of the literature addresses  one or the other  operation without discussing its combination with the other one.

All classic demosaicing methods have been proposed for noise free CFA images, while denoising algorithms have been designed for color or gray level images only considering additive white noise.  
Yet the input data is in reality different: it is either a CFA image with  noise, or a demosaiced image with structured noise.  Therefore, we can distinguish three main  pipeline strategies: denoising first followed by demosaicing ($\A\&\B$), demosaicing first followed by denoising ($\B\&\A$), and joint demosaicing-denoising.
It might be argued that with the advent of deep learning, the joint operation will become standard and the first two solutions obsolete. But there are three good reasons to address them. The first one is that, contrary to classic image processing chains, processing chains based on deep learning remain domain and device dependent. In other terms, even if they can give the best results on a given test set or device, there is not guarantee that they will deliver good results on out of domain images, or on new devices.
Hence, even with slightly apparent lower performance, classic algorithms still retain their value. Secondly, as has been verified many times, insight obtained by combining classic algorithms leads to conceive better deep learning structures. 
Last but not least, 
classical algorithms are characterized by computational efficiency and suitability for acceleration. This is exemplified by the successful implementation of classical algorithms, such as the BM3D algorithm, on select mobile devices, made possible through the adoption of advanced process chips, along with continued efforts in algorithmic enhancement and optimization. This accomplishment underscores the promising potential for classical algorithms to extend their reach to a broader spectrum of edge computing devices in the foreseeable future. In contrast, the computational demands of neural networks present challenges when it comes to deployment on low-performance hardware.
For these reasons, we shall focus here on a comparison of denoising first followed by demosaicing ($\A\&\B$) with demosaicing first followed by denoising ($\B\&\A$), and to generalizations of both approaches.

Currently, the most popular classic pipeline is the $\A\&\B$ scheme. This is determined by two basic assumptions. First, after demosaicing, the noise becomes correlated and no longer retains its independent identically distributed (i.i.d) white Gaussian properties. This  has a negative impact on traditional denoising algorithms that rely on additive white Gaussian noise (AWGN). Second, state-of-the-art demosaicing algorithms are often designed on a noise-free basis. As a result, many state-of-the-art works \cite{paliy2007demosaicing,park2009case,kalevo2002noise,zhang2009pca}
operate under the assumption that $\A\&\B$ outperforms $\B\&\A$. 

The advantage of $\A\&\B$ pipelines is that many excellent denoisers can be applied directly, such as model-based TV \cite{Rudin1992TV,Hintermuller2014adaptive,Chowdhury2020Poisson,Hu2020Spatial},  non-local~\cite{buades2005review,lebrun2013nonlocal,Jin2015nonlocal,jin2018convergence,jin2017nonlocal}, BM3D~\cite{Dabov2007BM3D,dabov2007color}, low rank~\cite{gu2014weighted,guo2022Gaussian} and deep learning-based methods~\cite{zhang2017beyond,Zhang2018FFDNet,Fang2021Multilevel,Yu2021Fast}, because the statistical nature of the noise is maintained. However, these methods are designed and optimized for grayscale or color images and need to be adapted for application to CFA images~\cite{park2009case,danielyan2009cross}. Meanwhile, demosaicing algorithms designed  on noise-free images can be applied directly after the noise is removed, e.g., \cite{hamilton1997adaptive,Wu2006Temporal,mairal2009non,pekkucuksen2010gradient,buades2011self,zhang2011color,kiku2013residual,Liang2013Wavelet,kiku2014minimized,kiku2016beyond,wu2016demosaicing,tan2017color,Jin2021IPOL}.

For example Park \emph{et al.}~\cite{park2009case} consider the classic Hamilton-Adams (HA)~\cite{hamilton1997adaptive} and a frequency-domain algorithm~\cite{dubois2005frequency} for demosaicing, combined with two denoising methods, BLS-GSM~\cite{portilla2003image} and CBM3D~\cite{dabov2007color}. 
This combination raises the question of adapting BM3D to a CFA. 
To do so, the authors first transform the noisy CFA image into the half-size 4-channel image formed by joining the four observed raw values (R,G,G,B) in each four pixel block, then remove noise channel by channel via BM3D \cite{Dabov2007BM3D}, finally get the denoised CFA image by the inverse color transform.  However,  this leads to a checkerboard effect that becomes more noticeable for higher noise levels. Similarly, BM3D-CFA~\cite{danielyan2009cross} removes noise directly from the CFA color array, which builds 3D blocks from the same CFA configuration.
BM3D-CFA was considered to be a systematic improvement method over~\cite{zhang2009pca}, in which the method~\cite{zhang2005color} was used as demosaicing method for their comparison of the result after demosaicing.  Analogously, \cite{chatterjee2011noise} adjusted  NL-means \cite{buades2005review} to the CFA image. Zhang \emph{et al.}~\cite{zhang2014joint} uses a filter \cite{alleysson2005linear} to extract the luminance of the CFA image. The authors of \cite{zhang2009pca} proposed a PCA-based CFA denoising method that makes full use of spatial and spectral correlation. In \cite{patil2016poisson}, Patil and Rajwade remove Poisson noise from CFA images using dictionary learning.

In general, the classical denoising algorithms (such as BM3D, NL-means) can all be adapted to accommodate CFA image denoising in the $\A\&\B$ strategy. 
Several of them~\cite{paliy2007demosaicing,park2009case,kalevo2002noise,zhang2009pca} address this realistic case  by processing the noisy CFA images as a  half-size 4-channel color image (with one red, two green and one blue channels) and then apply a multichannel denoising algorithm to it. 
Albeit  the $\A\&\B$ pipeline  maintains the independent and identically distributed property of the white Gaussian noise (Poisson noise can be transformed to Gaussian noise by the classical Anscombe transform \cite{Anscombe1948}), the disadvantage is the reduced resolution of the image (half size), which leads to loss of image detail after denoising.
Another issue is that it does not take advantage of the relative spatial position of the R, G, and B pixels due to the separation of the image into four independent channels (R,G,G,B) during denoising, resulting in the color distortion problem. Meanwhile, since G is separated into two independent G channels, the difference between the two G channels after denoising causes checkerboard artifacts.

The $\B\&\A$ pipeline was considered for better image detail preservation and to avoid checkerboard artifacts. Unfortunately, there is not many literatures on such pipelines. This is due to the strong spatial and chromatic correlation of the image noise after demosaicing. These correlations are generated by the demosaicing algorithm and are difficult to be modeled, which is detrimental to  model-based denoising algorithms. Condat made an attempt in~\cite{Condat2010simple}, where he first performed demosaicing and then projected the noise into the luminance channel of the reconstructed image and then denoised it based on the grayscale image. The idea was then further refined in \cite{Condat2012Joint,Condat2014Generic}.  This approach is similar to ours, but we will give a more elaborate theoretical explanation.

To avoid the accumulation of errors caused by the pipeline order, many researchers have proposed to perform a joint demosaicing and denoising~\cite{hirakawa2006joint,khashabi2014joint,gharbi2016deep}. With the popularity of deep learning, joint demosaicing denoising has gained great resolution and excellent performance. By constructing a large number of pairs of simulated data, joint demosaicing and denoising models can be readily trained. Algorithms based on convolutional neural networks (CNNs), such as \cite{syu2018learning}, exhibit performance far exceeding those of handcrafted algorithms \cite{monno2017adaptive}. After \cite{khashabi2014joint} introduced the first machine learning-based joint demosaicing and denoising method, Gharbi \emph{et al.}  \cite{gharbi2016deep} proposed the first deep learning model. Subsequently, a number of algorithms based on deep learning (such as \cite{dong2018joint,kokkinos2019iterative,elgendy9335264low,Liu_2020_CVPR,guo2023joint}) have been proposed. 
An unsupervised ``mosaic-to-mosaic'' training strategy for joint demosaicing and denoising was introduced by Ehret \emph{et al.} \cite{ehret2019joint}.
In \cite{guo9503334joint}, Guo \emph{et al.} focused on joint demosaicing and denoising of real-world burst images.
Further, Xing \emph{et al.} \cite{xing2021end} discussed end-to-end joint demosaicing, denoising and super-resolution. In the face of increasing network size and memory consumption, \cite{Guan2022Memory} proposed memory efficient joint demosaicing denoising for Ultra High Definition images.

The deep learning-based algorithms mentioned above achieve state-of-the-art performance, but suffer from a common problem of increasingly large network size and high computational complexity. This problem makes deploying these algorithms to devices, especially in low-power or portable devices, difficult to implement. 
Also, since deep learning algorithms rely on training, generalization issues might arise. For instance, if the noise range used during training is exceeded, or if the image is out of domain, the results might be significantly inferior to those obtained on a testing 
set. 
We have briefly summarized the advantages and drawbacks of the three pipelines in Table \ref{drawback}.

\begin{table}[t]
	\caption{Advantages and drawbacks of the three types of pipelines.
	}
	\label{drawback}
    \begin{center}
	\renewcommand{\arraystretch}{1.3} 
	\footnotesize
    \begin{tabular}{|c|m{2.8cm}<{\centering}|m{2.8cm}<{\centering}|m{2.8cm}<{\centering}|}
    \hline
     & $\A\&\B$ & $\B\&\A$ & Joint $\B\A$\\ \hline
    Advantages & The noise is maintained AWGN & Richer details & Better imaging quality \\ \hline
    Drawbacks & Detail loss and checkerboard artifacts & Spatial and chromaticity-related structural noise & High computational complexity and generalization concerns \\ \hline
    \end{tabular}
    \end{center}
\end{table}

In this paper, we address the problem of combining optimally and adapting  state-of-the-art  demosaicing and denoising algorithms.
A preliminary version of this study appeared in~\cite{jin2020Review}. There, we presented evidence showing that by demosaicking first and then denoising with a higher noise level (denoted $\B\&1.5\A$ schemes) yields substantially improved result compared with the classic configurations. This paper extends considerably that preliminary work. In particular, we conduct  thorough experiments and develop the arguments to confirm and to extend our conclusions. We first establish a model to optimize the denoising and demosaicing pipeline and use the black box optimizer CMA-ES~\cite{Hansen1996CMA-ES} to solve the optimization problem. The optimal results indicate that the $\B\&1.5\A$ scheme can get almost the same result as the CMA-ES optimum with a CPSNR value difference $\leq 0.08$ dB when $\sigma \leq 20$ and performs much better than $\A\&\B$ and $\B\&\A$ schemes. Then, we theoretically analyze the statistical properties of demosaiced noise and explain the reason why the $\B\&1.5\A$ scheme works well.
A series of experiments leads us to conclude  that  the $\B\&1.5\A$ scheme is always superior to the $\A\&\B$ and $\B\&\A$ ones.  
For large noise, the best scheme is more complex and has three stages, but we shall show that the $\B\&1.5\A$ scheme still is competitive. 
Our conclusions are different and actually opposite to  those of~\cite{paliy2007demosaicing,park2009case,kalevo2002noise,zhang2009pca}. 
The advantages of $\B\&1.5\A$ scheme seem to be linked to the fact that this scheme does not handle  half size 4-channels color image; it therefore uses the classic denoising methods directly on a full resolution color image;  
this results in more details being preserved and avoids checkerboard artifacts or loss of details.
These conclusions also provide theoretical support for real sRGB image denoising \cite{Guo2019Toward} which removes noise from full color images after demosaicing. 
The fact that $\B\&1.5\A$ schemes improve on the results of  raw image denoising will be verified by experiments carried out on two benchmarks, the Smartphone Image Denoising
Dataset (SIDD)~\cite{Abdelhamed2018} and the Darmstadt Noise Dataset (DND)~\cite{2017DND}.

The rest of this paper is structured as follows. 
In Section~\ref{sec:framework} we discuss how to apply demosaicing followed by denoising to CFA images. 
In Section~\ref{sec:pipeline}, the black box optimizer CMA-ES is  used to find the most general 3-step strategy. The results confirm the preference for $\B\&\A$ schemes in presence of moderate noise, and lead to a refinement for high noise levels with an $\A\&\B\&\A$ scheme.
In Section~\ref{analysisofBand15A}, we are led to define and analyze the statistical properties of the demosaicing residual noise in RGB and in a transformed  space that decorrelates the color channels. Then, using these statistical properties, we find experimentally the appropriate noise level that must be used for the  denoising method after demosaicing in a $\B\&\A$ scheme.   
Section~\ref{sec:experiment} compares our strategy with other state-of-the-art ones on simulated  and real image datasets. 
Section~\ref{sec:conclusion} concludes.

\section{The demosaicing and denoising pipeline}
\label{sec:framework}

The denoising and demosaicing pipeline consists in solving the ill-posed problem
\begin{equation}\label{model noisycfa}
    \mathbf{v}=\mathrm{Bayer}(\mathbf{u})+\epsilon,
\end{equation}
where $\mathbf{v}\in \mathbb{R}^{m\times n \times 3}$ is the observed noisy mosaicked image, $\mathrm{Bayer}$ is the Bayer color filter, $\mathbf{u}=(\mathbf{R},\mathbf{G},\mathbf{B})\in \mathbb{R}^{m\times n \times 3}$ is the latent ground truth color image and $\epsilon$ is Gaussian noise  with zero mean and standard deviation $\sigma$. 
As stated in the introduction, we will consider the problem of combining demosaicing and denoising, i.e. which one should be executed first? This brings us to two main pipelines: $\B\&\A$ (demosaicing then denoising),  $\A\&\B$ (denoising then demosaicing). In \cite{jin2020Review}, we reached the preliminary conclusion: demosaicing should be executed with higher priority and subsequent denoising needs to be adjusted.
In the next section we will propose to consolidate (and partly modify) this  conclusions by optimizing freely a 3-step procedure. Let $\sigma_{1}$ and $\sigma_{2}$ be the noise level hyperparameters of $\A\&\B$ and $\B\&\A$ respectively.

The restored image can be evaluated by subjective criteria such as visual quality and by objective criteria such as the color signal-to-noise ratio (CPSNR) \cite{alleysson2005linear},  defined by
\begin{equation}\label{EQ: psnr}
\mathrm{CPSNR}(\widehat{\mathbf{u}}) = 10 \log_{10} \frac{255^2}{ \mathrm{MSE} (\widehat{\mathbf{u}})}, 
\quad \text{with}\\
\end{equation}
\begin{equation*}
\mathrm{MSE}(\widehat{\mathbf{u}}) = \frac{1}{m\times n \times 3} \|\widehat{\mathbf{u}}-\mathbf{u}\|^{2}_{F},
\end{equation*}
where $\|\cdot\|_{F}$ is the Frobenius norm, $\mathbf{u}$ denotes the  ground truth image and  $\widehat{\mathbf{u}}$ is the estimated color image.

\begin{figure}[t]
	\begin{center}
		\renewcommand{\arraystretch}{0.7} \addtolength{\tabcolsep}{-4pt}
		{\fontsize{6pt}{\baselineskip}\selectfont
			\begin{tabular}{ccccc}
				&
				\includegraphics[width=0.15\linewidth]{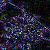}&
				\includegraphics[width=0.15\linewidth]{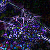}&
				\includegraphics[width=0.15\linewidth]{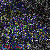}&
				\includegraphics[width=0.15\linewidth]{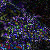} \\
				\includegraphics[width=0.15\linewidth]{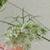}&
				\includegraphics[width=0.15\linewidth]{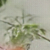}&
				\includegraphics[width=0.15\linewidth]{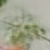}&
				\includegraphics[width=0.15\linewidth]{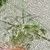}&
				\includegraphics[width=0.15\linewidth]{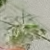}
				\\
				(a) Ground truth & (b) JCNN & (c) $\A\&\B$ & (d) $\B\&\A$ & (e) $\B\&1.5\A$   \\
				& ~~27.46 dB & ~~25.69 dB & ~~25.38 dB & ~~26.95 dB \\
			\end{tabular}
		}  
	\end{center}
	\caption{Image details at $\sigma = 20$. The lower row is the reconstructed image, and the upper row is the difference between the reconstructed image and ground truth. 
	$\A$: cfaBM3D or CBM3D denoising; $\B$: RCNN demosaicing. $1.5\A$ means that if the noise level is $\sigma$,  the input noise level parameter of denoising method $\A$ is $\sigma_2=1.5{{\sigma}}$.}
	\label{Fig: bookp4}
\end{figure}

Park {\em \emph{et al.}}~\cite{park2009case} argued that demosaicing
introduces chromatic and spatial correlations to the noise, which is no longer i.i.d. white Gaussian and therefore harder to model and eliminate.
In~\cite{kalevo2002noise} the authors argue that   $\A\&\B$  schemes with a proper parameter are more efficient than $\B\&\A$ schemes.
Figure~\ref{Fig: bookp4} (d) shows an example where a noisy CFA image with noise of standard deviation ${{\sigma}}$ was first  demosaiced by RCNN~\cite{tan2017color} and then restored by CBM3D~\cite{dabov2007color} assuming a noise parameter $\sigma_{2} = {{\sigma}}$. The output of CBM3D with $\sigma_{2}={{\sigma}}$ has a strong residual noise.
A similar behavior is also observed with other image denoising algorithms such as nlBayes~\cite{jin2020Review}.
Based on this argument several papers~\cite{park2009case,zhang2009pca,akiyama2015pseudo,lee2017denoising}  propose raw CFA denoising methods applicable before demosaicing.

\begin{figure}
	\begin{center}
		\renewcommand{\arraystretch}{0.5} \addtolength{\tabcolsep}{-4.5pt} {%
			\fontsize{8pt}{\baselineskip}\selectfont
			\includegraphics[width=0.8\linewidth]{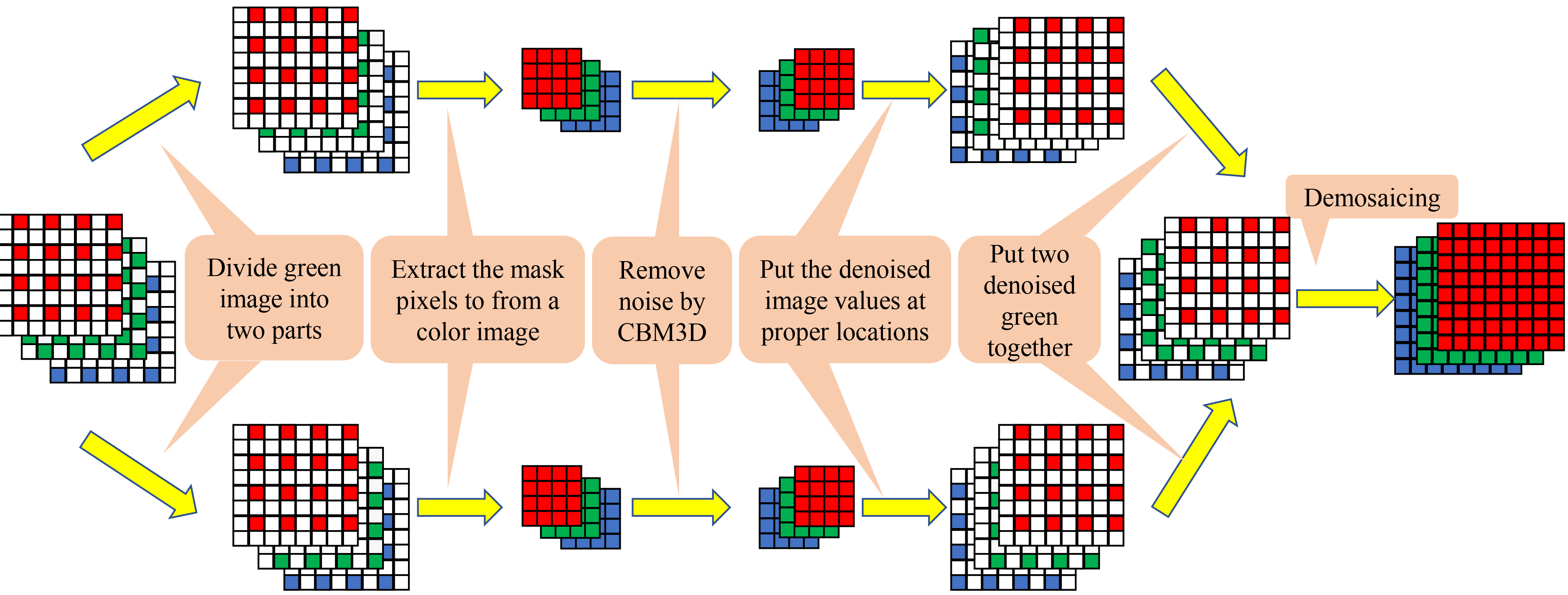}
		}
		\caption{
		The framework used for  denoising before demosaicing using an RGB denoiser. The Bayer CFA image is split in two half resolution RGB images, each one with a different green. Both RGB images are denoised independently. Then the pixels of both results are recombined into a denoised Bayer CFA image. The last step consists in applying a demosaicing algorithm.}
		\label{Fig framework}
	\end{center}
\end{figure}

Other denoising methods that are not explicitly designed to handle raw CFA images (such as CBM3D and nlBayes) can also be adapted  to noisy CFA images by rearranging the CFA image into a half-size four-channels image~\cite{park2009case}, or two half-size three-channel images as shown in Figure~\ref{Fig framework}.
In our comparative experiments, CBM3D will be used to process CFA images, which is the scheme in Figure~\ref{Fig framework}, we will denote this method as cfaBM3D.

In the case of splitting the raw image into two half-size 3-channel images (see Figure~\ref{Fig framework}), both images are  denoised independently and  the denoised pixels are recombined. 
Each half-size image contributes one green pixel to the denoised CFA image, while the red and blue pixels are averaged. 
Despite the $\A\&\B$ pipeline effectively eliminates noise, it  is not good at preserving details and produces artifacts such as checkerboard effect.
Indeed, due to the rearrangement of the CFA pixels, much image detail is lost in the image after applying an $\A\&\B$ scheme. 
In addition, this procedure introduces visible checkerboard artifacts for noise levels ${\sigma}>10$. 
These artifacts can be observed in Figure~\ref{Fig: bookp4}~(c).
To address this last issue, Danielyan \emph{et al.}~\cite{danielyan2009cross} proposed BM3D-CFA, which amounts to denoise four different mosaics of the same image before aggregating the four values obtained for each pixel. In practice, we observed that BM3D-CFA and the cfaBM3D method described above attain very similar results. The main difference between the two comes with the execution time, as for cfaBM3D a fast GPU implementation is available~\cite{Davy2021BM3D}. Depending on the experiment we will use one or the other.

Jin \emph{et al.}~\cite{jin2020Review} revised  the $\B\&\A$ pipeline and observed that a very simple modification of the noise parameter of the denoiser $\A$ coped with the structure of demosaiced noise, and led to  efficient denoising \textit{after} demosaicing, i.e. a $\B\&1.5\A$ pipeline.
This  allows for a better preservation of  fine structure often smoothed by the $\A\&\B$ schemes, and   checkerboard artifacts are  no longer observed (see Figure~\ref{Fig: bookp4}~(e)). 
In terms of  quality  and speed, demosaicing $\B$ can be done by a fast algorithm RCNN~\cite{tan2017color} followed by CBM3D denoising $1.5\A$, namely CBM3D applied with a noise parameter equal to $\sigma_{2} = 1.5{{\sigma}}$.

Figure~\ref{Fig: bookp4} also illustrates that $\A\&\B$ has better CPSNR than $\B\&\A$. However, the performance of $\B\&1.5\A$ pipeline is much superior to both $\B\&\A$ and $\A\&\B$. Is $\B\&1.5\A$ pipeline the optimal one? In Section~\ref{sec:pipeline}, we will explore a more generic optimal pipeline of denoising and demosaicing to confirm this optimality for moderate noise, and a near optimality for large noise. In Section~\ref{analysisofBand15A}, based on the analysis of demosaiced noise we shall seek an explanation of the efficiency of $\B\&1.5\A$.

\begin{figure}
	\begin{center}
			\includegraphics[width=0.8\linewidth]{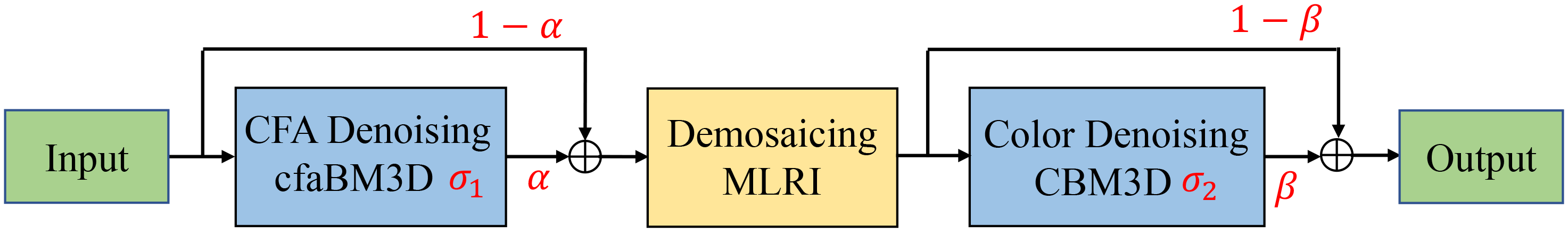}
		\caption{Generic raw image processing pipeline. This pipeline structure allows for an arbitrary order between $\A$ and $\B$ and sets free their parameters. We use the CMA-ES algorithm to optimize the parameter $\alpha$, $\beta$, $\sigma_1$, $\sigma_2$ in the pipeline.} 
		\label{cma_pipeline}
	\end{center}
\end{figure}

\section{Pipeline optimization and analysis}
\label{sec:pipeline}

\begin{figure*}[t]
		\centering
		\renewcommand{\arraystretch}{0.6} \addtolength{\tabcolsep}{-3pt} {%
			\fontsize{8pt}{\baselineskip}\selectfont
			\begin{tabular}{cccc}
			\includegraphics[width=0.23\textwidth]{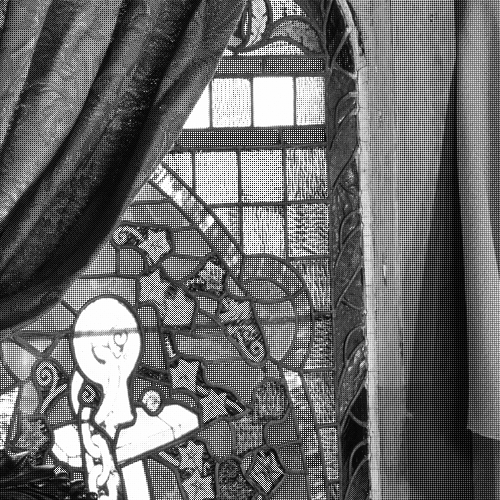} &
			\includegraphics[width=0.23\textwidth]{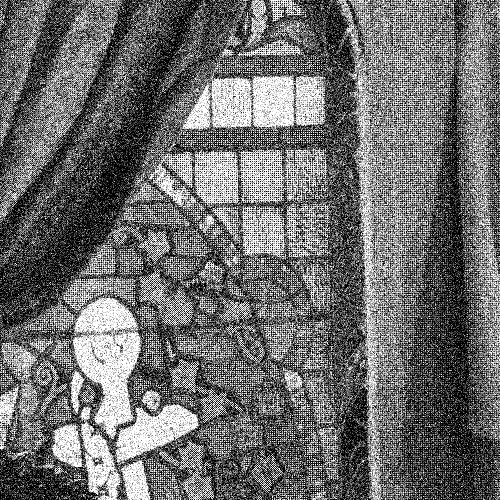} &
 		\includegraphics[width=0.23\textwidth]{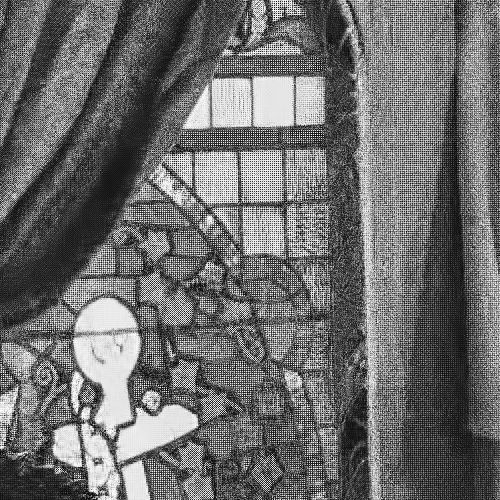} &
 		\includegraphics[width=0.23\textwidth]{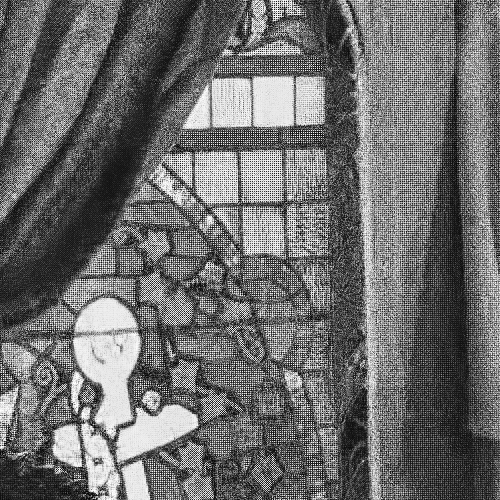} \\

            GT CFA image & Noisy CFA image & CFA Denoising & $\alpha$ linear combination \\

			\includegraphics[width=0.23\textwidth]{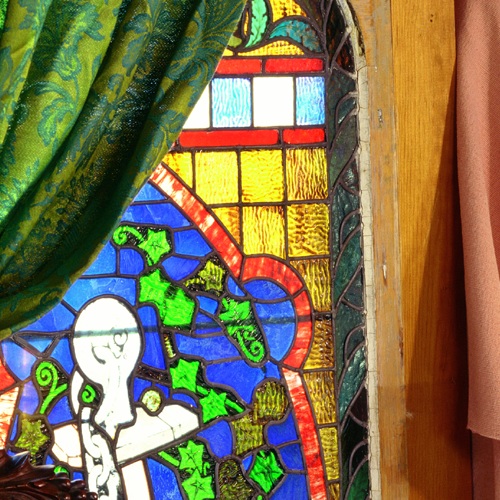} &
			\includegraphics[width=0.23\textwidth]{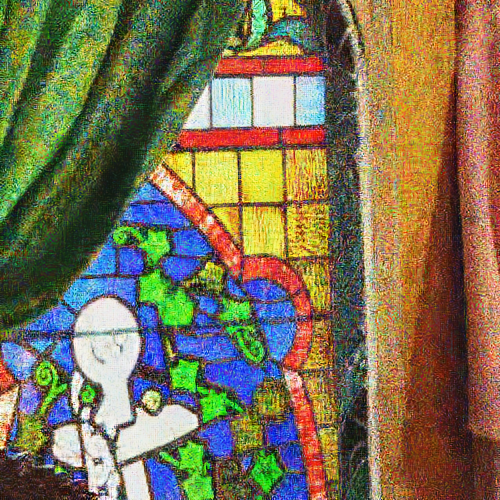} &
 		\includegraphics[width=0.23\textwidth]{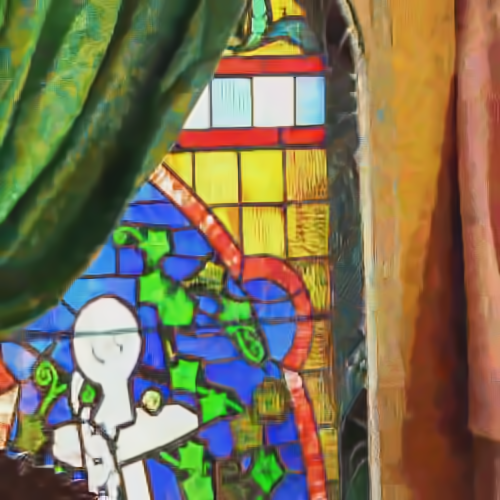} &
 		\includegraphics[width=0.23\textwidth]{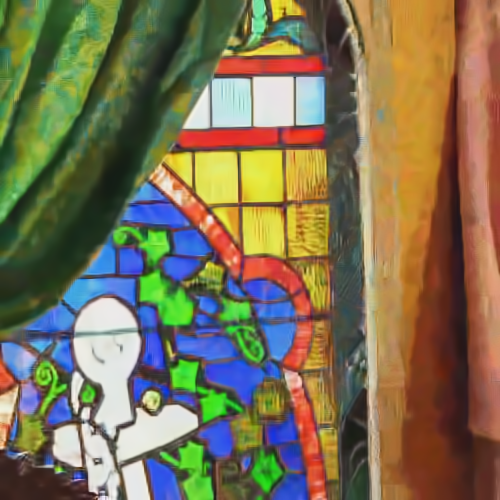} \\

			GT color image & Demosaicing & Color Denoising & $\beta$ linear combination \\
			\end{tabular}
		}
		\caption{A visual representation of the process in Figure \ref{cma_pipeline}, where the noise level is $\sigma = 60$. The parameter are $\alpha= 0.90$, $\beta= 0.99$, $\sigma_1= 34.50$, $\sigma_2= 54.42$. Since $\beta$ is always close to 1 in the pipeline, the visual difference between Color Denoising and the $\beta$ linear combination is not significant.}  
  
		\label{fig_cma_step}
\end{figure*}

In order to arrive at a rigorous decision in a more general framework,  we designed a generic $\A_{1}\&\B\&\A_{2}$ pipeline. The structure of the pipeline is illustrated in Figure~\ref{cma_pipeline}. This pipeline allows for an arbitrary order between $\A$ and $\B$ and sets free their parameters. It has two denoisers and four hyperparameters. The two denoisers are a CFA denoiser $\A_{1}$ (see Figure~\ref{Fig framework}) and a full color image denoiser $\A_{2}$, which respectively remove noise before  and after demosaicing. The four hyperparameters are $\alpha$ (that controls the weight of CFA denoising), $\beta$ (that controls the weight of color denoising), $\sigma_1$ (the noise standard deviation of the CFA denoiser), $\sigma_2$ (the noise standard deviation of the color denoiser). 
The results of the pipeline are visualised in Figure \ref{fig_cma_step}.
The final result of the pipeline is given by
\begin{equation} \label{EQ: final resultpipeline}
    \widehat{\mathbf{u}}=\beta\A_{2} (\B(\widetilde{\mathbf{v} }),\sigma_{2})+(1-\beta)\B(\widetilde{\mathbf{v} }),
\end{equation}
where
\begin{equation*}
    \widetilde{\mathbf{v} }=\alpha \A_{1}(\mathbf{v},\sigma_{1}) + (1-\alpha)\mathbf{v}.
\end{equation*}
It follows that if $\alpha=1$, $\beta=0$, $\sigma_{1}=\sigma$ and $\sigma_{2}=0$,  then $\widetilde{\mathbf{v}}=\A(\mathbf{v})$ and $\widehat{\mathbf{u}}=\B (\A(\mathbf{v}))$, i.e. the pipeline is  $\A\&\B$;
if $\alpha=0$, $\beta=1$, $\sigma_{1}=0$ and $\sigma_{2}=\sigma$, then $\widetilde{\mathbf{v}}=\mathbf{v}$ and $\widehat{\mathbf{u}}=\A (\B (\mathbf{v}))$, i.e. the pipeline is  $\B\&\A$; if $\alpha=0$, $\beta=1$, $\sigma_{1}=0$ and $\sigma_{2}=1.5\sigma$, then $\widetilde{\mathbf{v}}=\mathbf{v}$ and $\widehat{\mathbf{u}}=\A (\B (\mathbf{v}))$, i.e. the pipeline is  $\B\&1.5\A$
\cite{jin2020Review}.

\begin{figure*}[t]
		\centering
		\renewcommand{\arraystretch}{0.4} \addtolength{\tabcolsep}{-5pt} {%
			\fontsize{8pt}{\baselineskip}\selectfont
			\begin{tabular}{cccc}
			\multirow{1}{*}[50pt]{$\sigma=5$} &
			\includegraphics[width=0.30\textwidth]{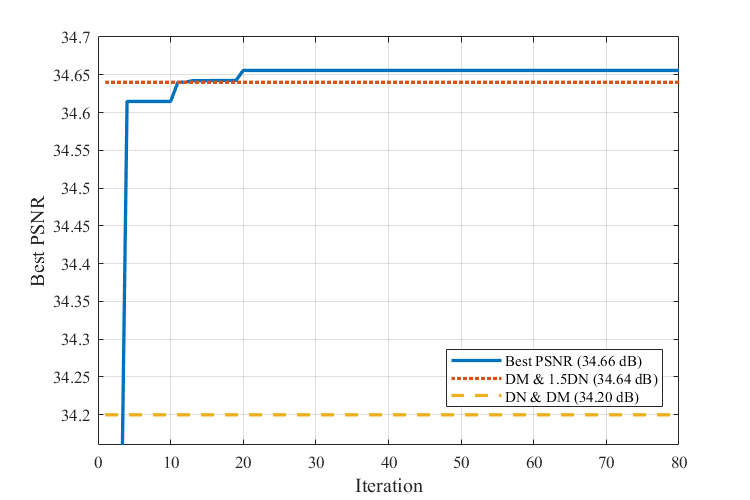} &
 			\includegraphics[width=0.30\textwidth]{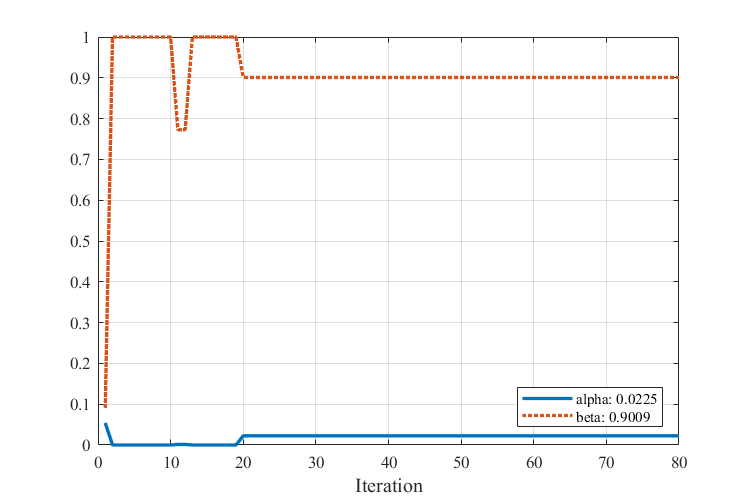} &
 			\includegraphics[width=0.30\textwidth]{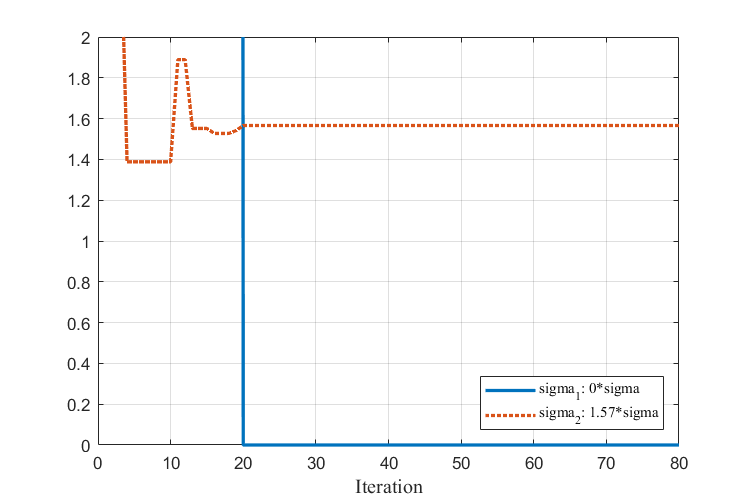} \\

 			\multirow{1}{*}[50pt]{$\sigma=20$} &
			\includegraphics[width=0.30\textwidth]{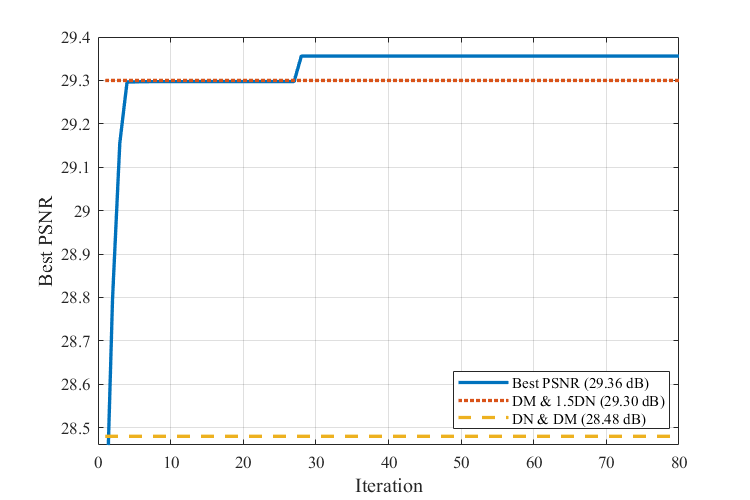} &
			\includegraphics[width=0.30\textwidth]{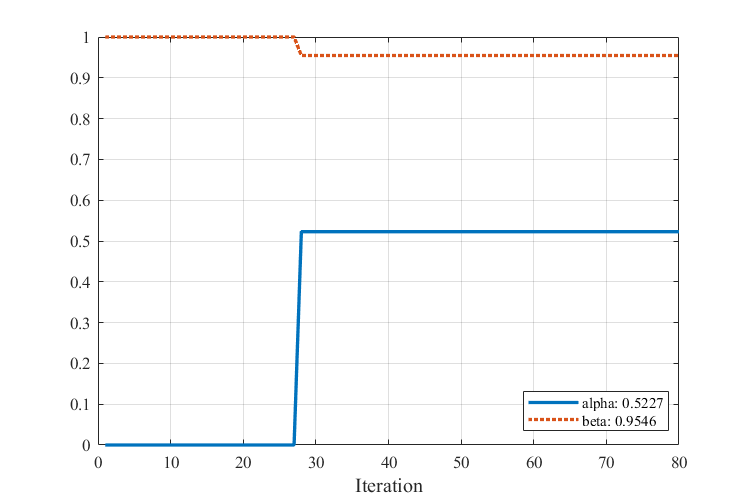} &
			\includegraphics[width=0.30\textwidth]{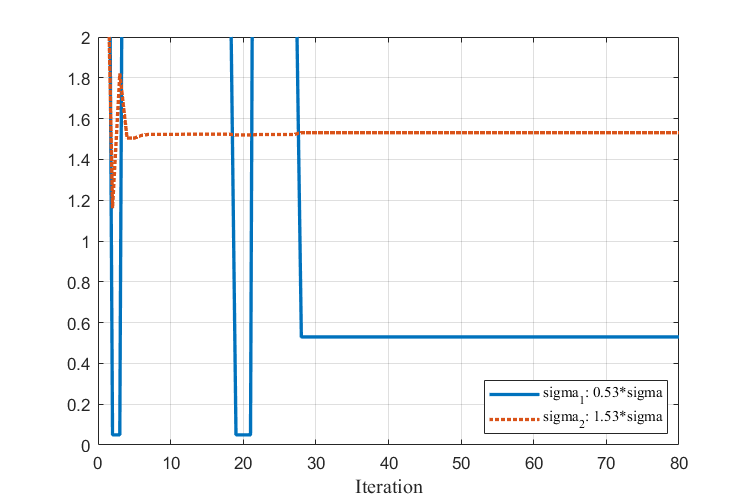} \\
  
 			\multirow{1}{*}[50pt]{$\sigma=60$} &
 			\includegraphics[width=0.30\textwidth]{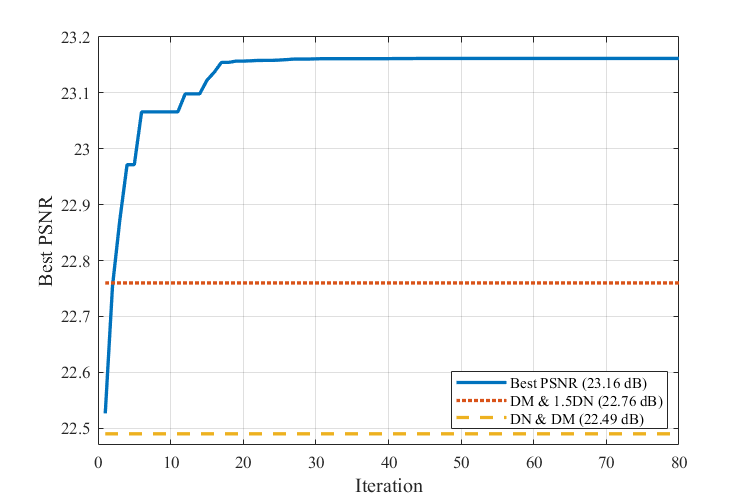} &
 			\includegraphics[width=0.30\textwidth]{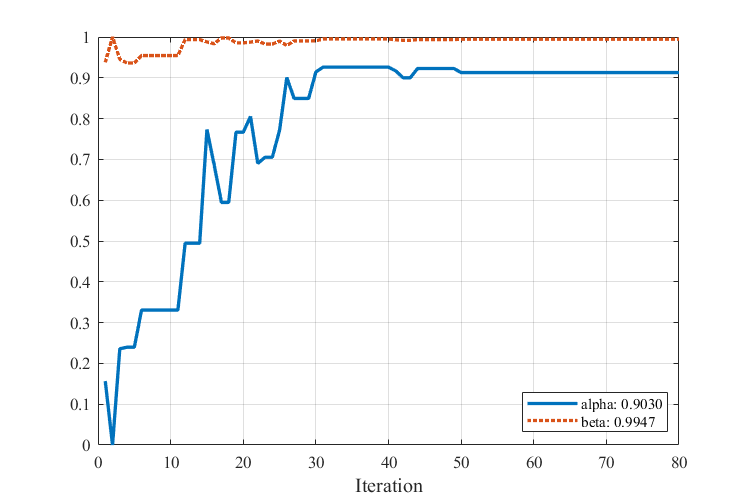} &
			\includegraphics[width=0.30\textwidth]{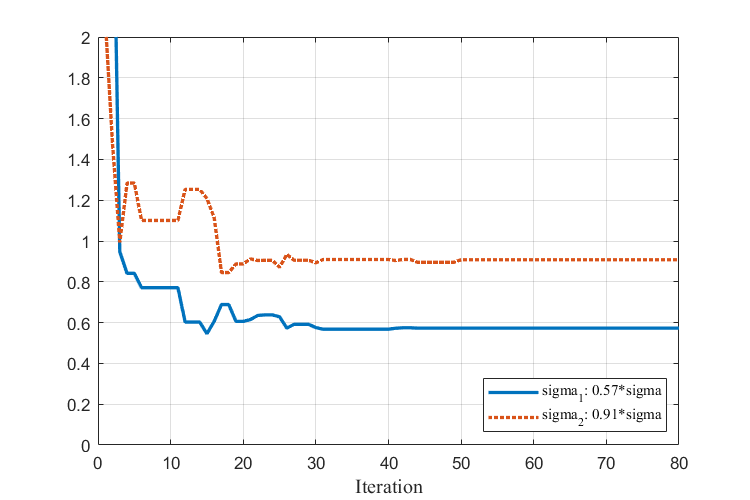} \\
			
			& (a) CPSNR & (b) $\alpha$ and $\beta$ & (c) $\sigma_1$ and $\sigma_2$
			\end{tabular}
		}
		\caption{Evolution of the result of iterating  CMA-ES when optimizing the  parameters $\alpha, \beta, \sigma_1, \sigma_2$ of the processing pipeline.}
		\label{fig_cma}
\end{figure*}

Our purpose is, for every noise level $\sigma$, to find  the optimal values $\{\alpha^*,\beta^*,\sigma_{1}^*,\sigma_{2}^*\}$ satisfying
\begin{equation}\label{EQ: optimal}
    \{\alpha^*,\beta^*,\sigma_{1}^*,\sigma_{2}^*\} =\mathop{\arg \max}_{ \{\alpha,\beta,\sigma_{1},\sigma_{2}\}} \mathrm{CPSNR}(\widehat{\mathbf{u}}),
\end{equation}
where $\widehat{\mathbf{u}}$ is defined by \eqref{EQ: final resultpipeline} and  CPSNR is defined in \eqref{EQ: psnr}.

\begin{table}[t]
	\caption{The optimization result of CMA-ES for the pipeline $\A_{1}\&\B\&\A_{2}$ (see Eq.~\eqref{EQ: final resultpipeline}), where $\sigma, \sigma_1, \sigma_2 \in [0, 255]$ and $\alpha, \beta \in [0, 1]$.  In this experiment $\B$ is always MLRI and $\A$ is CBM3D or cfaBM3D depending on the input data. }
	\label{table_cma}
    \begin{center}
    \renewcommand{\arraystretch}{1.1} \addtolength{\tabcolsep}{-1pt} 
	\footnotesize
    \begin{tabular}{clcccccc}
    \hline
        \multirow{2}{*}{$\sigma$} & \multirow{2}{*}{Method}  & \multirow{2}{*}{$\alpha$}  & \multirow{2}{*}{$\beta$} & \multirow{2}{*}{$\sigma_1$} & \multirow{2}{*}{$\sigma_2$} & CPSNR & CPSNR \\ 
       & & & & & & Imax & Kodak \\ \hline
       \multirow{4}{*}{5}  & $\A\&\B$    & 1.00 & 0.00 & 5.00 & 0 & 34.20 & 35.08 \\
                           & $\B\&\A$    & 0.00 & 1.00 & 0 & 5.00 & 34.18 & 35.03 \\
                           & $\B\&1.5\A$ & 0.00 & 1.00 & 0 & 7.50 & 34.64 & 35.77 \\
                           & CMA-ES 
                           & 0.02 & 0.90 & 0 & 7.83 & \textbf{34.66} & \textbf{35.78}\\ \hline
       \multirow{4}{*}{10} & $\A\&\B$    & 1.00 & 0.00 & 10.00 & 0 & 31.68 &    32.15 \\
                           & $\B\&\A$    & 0.00 & 1.00 & 0 & 10.00 & 31.55 & 31.62 \\
                           & $\B\&1.5\A$ & 0.00 & 1.00 & 0 & 15.00 & 32.35 & 32.99 \\
                           & CMA-ES 
                           & 0.51 & 0.92 & 6.81 & 12.98 & \textbf{32.43} & \textbf{33.02}\\ \hline
       \multirow{4}{*}{20} & $\A\&\B$    & 1.00 & 0.00 & 20.00 & 0 & 28.48 & 28.91 \\
                           & $\B\&\A$    & 0.00 & 1.00 & 0 & 20.00 & 28.07 & 27.75 \\
                           & $\B\&1.5\A$ & 0.00 & 1.00 & 0 & 30.00 & 29.30 & 29.85 \\
                           & CMA-ES 
                           & 0.52 & 0.95 & 10.58 & 30.63 & \textbf{29.36} & \textbf{29.91}\\ \hline
       \multirow{4}{*}{40} & $\A\&\B$    & 1.00 & 0.00 & 40.00 & 0 & 24.90 & 25.84 \\
                           & $\B\&\A$    & 0.00 & 1.00 & 0 & 40.00 & 24.16 & 24.05 \\
                           & $\B\&1.5\A$ & 0.00 & 1.00 & 0 & 60.00 & 25.46 & 26.53 \\
                           & CMA-ES 
                           & 0.82 & 0.98 & 23.46 & 41.79 & \textbf{25.74} & \textbf{26.72}\\ \hline  
       \multirow{4}{*}{50} & $\A\&\B$    & 1.00 & 0.00 & 50.00 & 0 & 23.62 & 24.83 \\
                           & $\B\&\A$    & 0.00 & 1.00 & 0 & 50.00 & 22.87 & 23.00 \\
                           & $\B\&1.5\A$ & 0.00 & 1.00 & 0 & 75.00 & 24.01 & 25.33 \\
                           & CMA-ES 
                           & 0.72 & 1.00 & 30.55 & 49.75 & \textbf{24.36} & \textbf{25.61}\\ \hline     
       \multirow{4}{*}{60} & $\A\&\B$    & 1.00 & 0.00 & 60.00 & 0 & 22.49 & 23.90 \\
                           & $\B\&\A$    & 0.00 & 1.00 & 0 & 60.00 & 21.83 & 22.24 \\
                           & $\B\&1.5\A$ & 0.00 & 1.00 & 0 & 90.00 & 22.76 & 24.26 \\
                           & CMA-ES 
                           & 0.90 & 0.99 & 34.50 & 54.42 & \textbf{23.16} & \textbf{24.60}\\ \hline     
    \end{tabular}
    \end{center}
\end{table}

Obviously, problem \eqref{EQ: optimal} is non-linear, non-convex and the gradients are not readily available. 
{In order to obtain the optimal solution of~\eqref{EQ: optimal} (and inspired by~\cite{mosleh2020hardware})}, we used the black box optimizer CMA-ES~\cite{Hansen1996CMA-ES}, which is a random search optimizer that is based on evolutionary strategies.
Unlike  common gradient optimization, CMA-ES does not  compute the gradient of the objective function. Only the ranking between candidate solutions is exploited for learning the sample distribution; neither derivatives nor even the function values themselves are required by the method~\cite{cmaes-matlab}.

We carried out experiments with different noise levels ($\sigma=5, 10, 20, 40, 50, 60$) on the images from the  Imax~\cite{zhang2011color} and Kodak~\cite{franzen1999kodak} datasets. In this  experiment we used the denoiser with the framework Figure~\ref{Fig framework}  
for $\A_1$, MLRI for $\B$ and CBM3D~\cite{dabov2007color} for $\A_2$. For each experiment,  $\{\alpha,\beta,\sigma_{1},\sigma_{2}\}$ were initialized randomly. Figure~\ref{fig_cma} illustrates the evolution of the CPSNR during the optimization with respect to $\{\alpha,\beta,\sigma_{1},\sigma_{2}\}$. In all cases, the parameters and the CPSNR stabilize after about $60$-iterations.  
The  final results are shown in Table~\ref{table_cma} along with results of the $\A\&\B $ method (cfaBM3D+MLRI)\footnote{Here the CFA image is divided into two half-size RGB images  then the noise is removed by CBM3D (see Figure~\ref{Fig framework}).}, the $\B\&\A$ method (MLRI+CBM3D) and $\B\&1.5\A$  (MLRI+1.5CBM3D as in \cite{jin2020Review}). 
When $\sigma \leq 20$  the optimal CMA-ES result is almost identical to the one of  $\B\&1.5\A$, and much better than $\A\&\B $ and $\B\&\A$. When $\sigma\geq 40$ the optimal CMA-ES result is much better than the ones obtained by   $\A\&\B $, $\B\&\A$ and $\B\&1.5 \A$. 
When $\sigma=5$, we observe that $\sigma_{1}=0$, which means that the pipeline is exactly $\B\&\A$ with parameter $\sigma_{2}/\sigma = 1.566$, i.e. $\B\&1.566\A$. When $\sigma\geq 10$, $\sigma_{1}$ is almost equal to $0.5\sigma$, however the CPSNR gain is only marginal.
The value $\sigma_{2}/\sigma$ decreases as $\sigma$ increases. Furthermore, from $\sigma=5$ to $60$, $\alpha$ increases from $0.0225$ to $0.9030$ while $\beta$ remains always larger than $0.9$. 
This means that applying denoising before demosaicing is not  important for low  noise levels, but becomes necessary when $\sigma$ increases, while applying denoising after demosaicing is always favorable, but with  a little smaller  denoising parameters.

When the noise level is high, the CPSNR of $\B\&1.5\A$ is 0.3 to 0.4 dB below the optimal value obtained by the $\A_1\&\B\&\A_2$ pipeline. however,  this requires almost doubling the computational complexity due to denoising.
Therefore, by trading-off image quality and computational cost, the simplified $\B\&1.5\A$ pipeline remains a good option and it is almost optimal for moderate noise. For this reason, we shall explore in detail this pipeline and the reasons of its near optimality in the next section.

\begin{figure}[t]
	\begin{center}
		\renewcommand{\arraystretch}{0.4} \addtolength{\tabcolsep}{-5.5pt} {%
			\fontsize{8pt}{\baselineskip}\selectfont
			\begin{tabular}{ccccc}
				\includegraphics[width=0.15\linewidth]{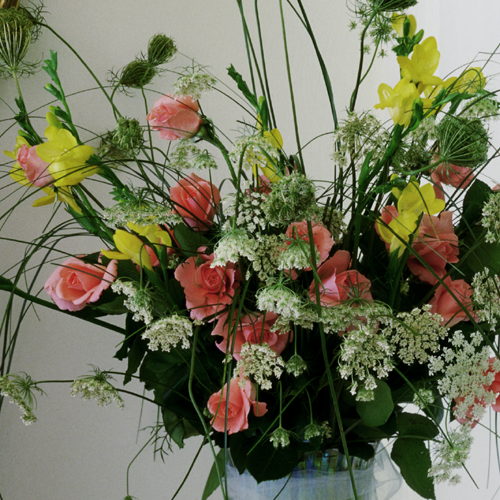}&
				\includegraphics[width=0.15\linewidth]{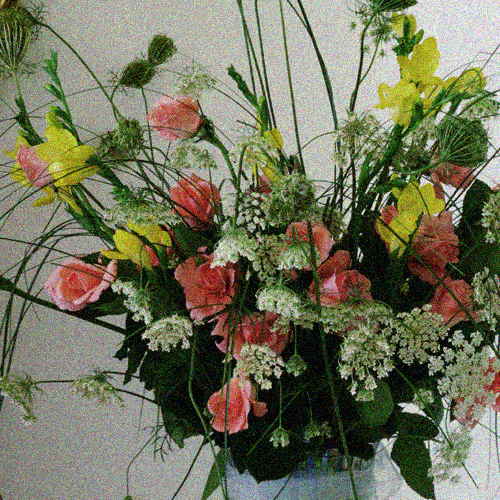}&
				\includegraphics[width=0.15\linewidth]{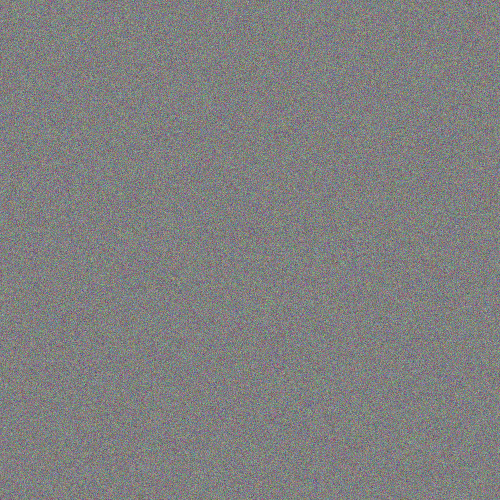}&
				\includegraphics[width=0.15\linewidth]{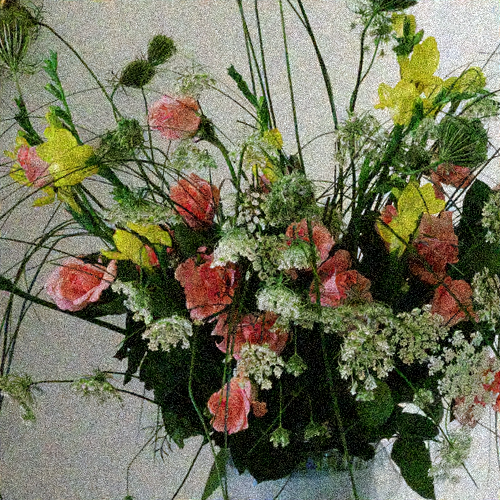}&
				\includegraphics[width=0.15\linewidth]{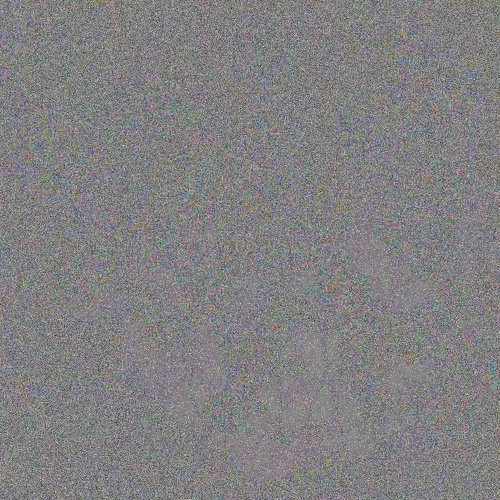}
				\\
				\includegraphics[width=0.15\linewidth]{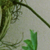}&
				\includegraphics[width=0.15\linewidth]{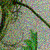}&
				\includegraphics[width=0.15\linewidth]{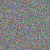}&
				\includegraphics[width=0.15\linewidth]{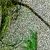}&
				\includegraphics[width=0.15\linewidth]{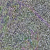}
				\\
				\includegraphics[width=0.15\linewidth]{./Fig_3P4.png}&
				\includegraphics[width=0.15\linewidth]{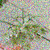}&
				\includegraphics[width=0.15\linewidth]{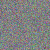}&
				\includegraphics[width=0.15\linewidth]{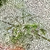}&
				\includegraphics[width=0.15\linewidth]{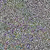}
				\\
				(a) Ground &  (b) AWGN  &  (c) AWGN  &  (d) RCNN & (e) RCNN  \\
				
				~~truth  & ~~image & ~~noise & ~~~image & ~~noise \\
			\end{tabular}
		} 
	\end{center}
	\caption{First row: (a) Ground truth Imax 3, (b) its noisy version, (c) added white noise (${{\sigma}} =20$), (d) demosaiced version of (b) by RCNN, (e) the {\it demosaiced noise}, namely the difference (d)-(a).
	Second and third rows: $50\times 50$ extracts from the first row.}
	\label{Fig: noisyimages}
\end{figure}

\section{Analysis of \texorpdfstring{$\B\&1.5\A$}{}}
\label{analysisofBand15A}

As we saw in Section~\ref{sec:pipeline},
The result of the $\B\& 1.5\A$ pipeline is almost equal to the result of the optimal $\A_1\&\B\&\A_2$ pipeline and much better than the $\A\&\B$ pipeline for all noise levels.    
The fact that a $\B\&1.5\A$ pipeline surpasses than a $\A\&\B$ scheme is surprising, considering that after demosaicing the noise is no longer white. Indeed, chromatic and spatial correlations have been introduced by the demosaicing, while the applied denoiser was conceived for white noise.   This apparent paradox leads us to analyze the behavior of demosaiced noise.

\begin{definition}
	Consider a ground truth color image
	 $( \mathbf{R}, \mathbf{G}, \mathbf{B})$ and its mosaic obtained by keeping  only one value of either 
	$ \mathbf{R}, \mathbf{G}, \mathbf{B}$ at each pixel, on a fixed Bayer  pattern. 
	Assume that white noise with standard deviation 
	${{\sigma}}$ 
	has been added to the mosaicked image, and that the resulting noisy mosaic has been demosaiced by $\B$, hence giving a noisy image $(\tilde  {\mathbf{R}}, \tilde  {\mathbf{G}}, \tilde  {\mathbf{B}})$. 
	We then call {\rm demosaiced noise} the difference 
	$(\tilde {\mathbf{R}}- \mathbf{R}, \tilde  {\mathbf{G}}- \mathbf{G}, \tilde  { \mathbf{B}}- \mathbf{B})$.  
\end{definition}	

Figure~\ref{Fig: noisyimages} illustrates the above definition. The \textit{demosaiced noise}	is nothing but the difference between the
demosaiced version of a noisy image and its underlying ground truth. 
The demosaiced  noise of column (d) is (visually) not significantly higher than  the white noise of column (b), but it is clearly no longer white, due to the introduction of chromaticity and spatial correlations. The properties of the \textit{demosaiced noise} depend on the demosaicing algorithm, as developed  in~\cite{jin2020Review}.  This paper compares $\B\&1.5\A$ pipelines composed of seven different state-of-the-art demosaicing algorithms (such as HA~\cite{hamilton1997adaptive}, GBTF~\cite{pekkucuksen2010gradient}, RI~\cite{kiku2013residual} and so on).
To understand empirically the right noise model to adopt after demosaicing, and following the conclusions of~\cite{jin2020Review}, we applied CBM3D after demosaicing with a noise parameter $\sigma_{2}$ corresponding to $\sigma$ multiplied by $(1.0,1.1, \cdots, 1.9)$. These experiments show that the optimal parameter interval is $[1.4, 1.7]$ and that  the optimal factor is 1.5.

\begin{table}[t]
	\caption{RMSE between ground truth and demosaicked image for different demosaicking algorithms in presence of noise of standard deviation $\sigma$. }
	\label{Table rmse}
	\begin{center}
		\renewcommand{\arraystretch}{1.25} \addtolength{\tabcolsep}{0pt} \vskip3mm
		\small
		\rowcolors{2}{gray!25}{white}
		\begin{tabular}{c|ccccc}
			\rowcolor{gray!50}
			\hline
			${{\sigma}}$& HA & GBTF & RI & MLRI & RCNN \\	\hline
			$1$&5.04&  5.10&  4.17&  4.06&  3.21\\
			$3$&5.70&  5.79&  4.97&  4.88&  4.17\\
			$5$&6.78&  6.87&  6.12&  6.10&  5.59\\
			$10$&10.18& 10.27&  9.53&  9.74&  9.65\\
			$15$&13.93& 14.01& 13.15& 13.64& 13.87\\
			$20$&17.75& 17.83& 16.77& 17.56& 18.04\\
			$30$&25.36& 25.42& 23.94& 25.30& 26.21\\
			$40$&32.67& 32.76& 30.77& 32.64& 33.98\\
			$50$&39.58& 39.71& 37.25& 39.55& 41.21\\
			$60$&46.14& 46.35& 43.43& 46.11& 47.95\\
			\hline
		\end{tabular}
	\end{center}
\end{table}

This surprising result would seem to imply that demosaicing increases noise. But this is not the case, as illustrated in Table~\ref{Table rmse}, which gives the noise standard deviation estimated as the mean RMSE of the demosaiced images from the Imax~\cite{zhang2011color} dataset with different noise levels. For low noise  ($\sigma=1$) the large demosaicing error of about 4 clearly is caused by the demosaicing itself. 
However, for $\sigma>10$  the RMSE of the demosaiced image tends to be roughly equal to 3/4 of the initial  noise standard deviation. In short, as expected from an interpolation algorithm, demosaicing (slightly) decreases the noise standard deviation. 
This is also consistent with the visual results observed in Figure~\ref{Fig: noisyimages}.

\begin{figure}[t]
	\begin{center}
		\renewcommand{\arraystretch}{0.2} \addtolength{\tabcolsep}{-5.5pt} 
		{%
			\small
			\begin{tabular}{cccc}
				\includegraphics[width=0.2\linewidth]{./Fig_NoisyImg20I3P4.png}&
				\includegraphics[width=0.2\linewidth]{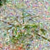}&
				\includegraphics[width=0.2\linewidth]{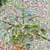}&
				\includegraphics[width=0.2\linewidth]{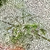}
				\\
				\includegraphics[width=0.2\linewidth]{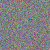}&
				\includegraphics[width=0.2\linewidth]{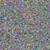}&
				\includegraphics[width=0.2\linewidth]{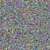}&
				\includegraphics[width=0.2\linewidth]{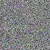}
				\\
				\includegraphics[width=0.2\linewidth]{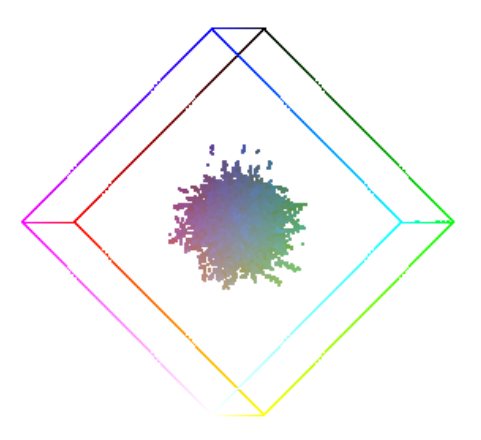}&
				\includegraphics[width=0.2\linewidth]{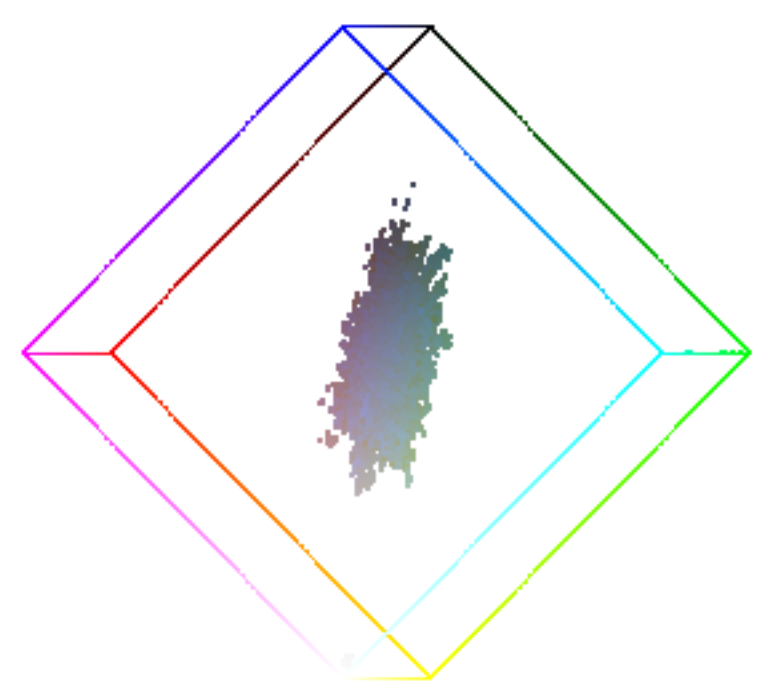}&
				\includegraphics[width=0.2\linewidth]{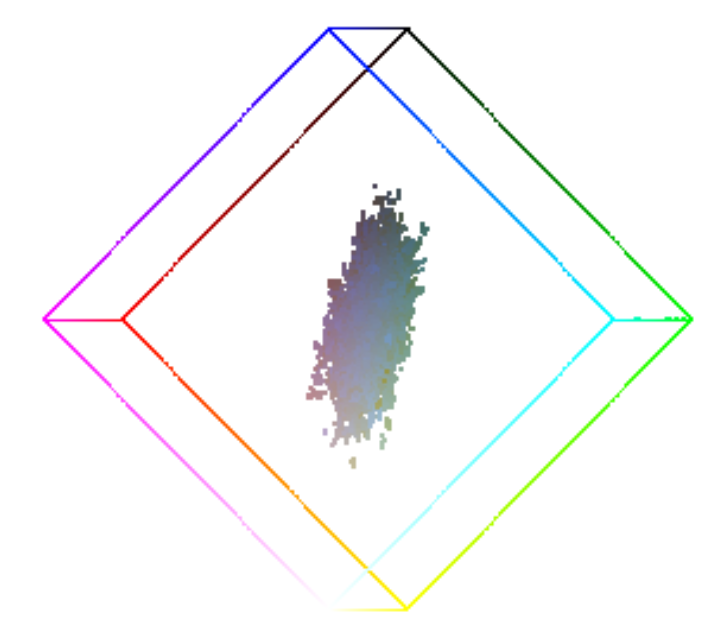}&
				\includegraphics[width=0.2\linewidth]{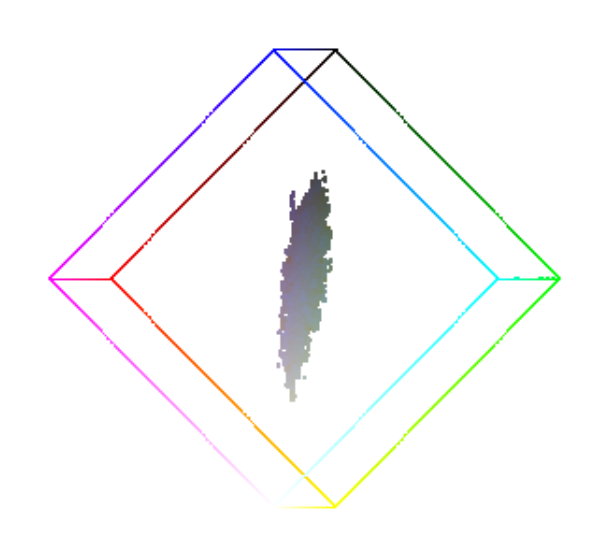}\\
				(a) AWGN  & (b)  HA   &  (c)  MLRI   &  (d)  RCNN 
			\end{tabular}
		} 
	\end{center}
	\caption{AWGN image and demosaicing noise with standard deviation $\sigma = 20$ for respectively HA, MLRI, RCNN. Last row: the color spaces (in standard (R,G,B) Cartesian coordinates) of each noise, presented in their projection with maximal area.  As expected, the AWG color space is isotropic, while the color space after demosaicing is elongated in the luminance  direction $\mathbf{Y}$  and squeezed in the others.  This amounts to an increased noise standard deviation for $\mathbf{Y}$ after demosaicing, and less noise in the chromatic directions.  See table Table \ref{Table Correlationpixels4level20} for quantitative results.}
	\label{Fig: noisyimageskind20}
\end{figure}

At first sight, this $3/4$ factor  contradicts the observation that denoising with a parameter $\sigma_{2}=1.5\sigma$  yields better results. This leads us to further analyze the structure of the demosaiced residual noise.
To that aim, we applied an orthonormal Karhunen-Loeve transform to the residual noise to maximally decorrelate the color channels \cite{malvar2004high, 1980Color}. This transform is commonly used in denoising algorithms~\cite{2012Secrets} such as CBM3D~\cite{dabov2007color}. Here, we used a transform $(\mathbf{R},\mathbf{G},\mathbf{B}) \to (\mathbf{Y},\mathbf{C}_{1},\mathbf{C}_{2})$, in which the luminance direction is $\mathbf{Y}= \frac{\mathbf{R}+\mathbf{G}+\mathbf{B}}{\sqrt3}$ and the orthogonal vectors $\mathbf{C}_{1}$ and $\mathbf{C}_{2}$ are arbitrarily chosen as in [45], which is defined as
\begin{align}
\left(
  \begin{array}{ccc}
     \mathbf{Y}  \\
     \mathbf{C_1} \\
     \mathbf{C_2} \\
  \end{array}
\right)=
\left(
  \begin{array}{ccc}
   1/\sqrt{3} & 1/\sqrt{3} & 1/\sqrt{3} \\
    1/\sqrt{2} &0  & - 1/\sqrt{2}\\
   1/\sqrt{6}  & -2/\sqrt{6} & 1/\sqrt{6} \\
  \end{array}
\right)\left(
  \begin{array}{ccc}
     \mathbf{R}  \\
     \mathbf{G} \\
     \mathbf{B} \\
  \end{array}
\right).
\end{align}

\begin{table}[t]
	\caption{Noise intensity. Variance and covariance of $(\mathbf{R},\mathbf{G},\mathbf{B})$ and $(\mathbf{Y},\mathbf{C}_{1},\mathbf{C}_{2})$ between pixels $(i,j)$ and $(i+s,j+t)$, $s,t = 0,1,2$ first for AWGN (a) with standard deviation $\sigma = 20$, then for its demosaiced versions by HA (b), RI (c),  MLRI (d) and RCNN (e). }
	\label{Table Correlationpixels4level20}
	\begin{center}
		\renewcommand{\arraystretch}{1} \addtolength{\tabcolsep}{-4.8pt}
		\footnotesize
		\begin{tabular}{l |rrrrrrrrrrr}
			\rowcolor{gray!50}
			
			& \scriptsize (\emph{i,j}) & \scriptsize (\emph{i,j}+1) &  \scriptsize (\emph{i,j}+2) &  \scriptsize (\emph{i}+1,\emph{j}) & \scriptsize (\emph{i}+1,\emph{j}+1)&  \scriptsize (\emph{i}+1,\emph{j}+2)  &  \scriptsize (\emph{i}+2,\emph{j})  &  \scriptsize (\emph{i}+2,\emph{j}+1) &  \scriptsize (\emph{i}+2,\emph{j}+2)
			\\\hline\hline
			\footnotesize R&\g{400.6}&\g{0.6}&\g{0.4}&\g{0.7}&\g{0.1}&\g{0.7}&\g{0.3}&\g{0.2}&\g{0.8}\\
			\footnotesize G&\g{401.7}&\g{0.5}&\g{1.1}&\g{0.1}&\g{0.3}&\g{0.9}&\g{1.0}&\g{0.6}&\g{0.4}\\
			\footnotesize B&\g{400.2}&\g{1.2}&\g{0.1}&\g{0.5}&\g{0.6}&\g{0.0}&\g{1.9}&\g{0.3}&\g{1.9}\\
			\hline\hline
			\footnotesize Y&\g{399.6}&\g{1.1}&\g{0.1}&\g{0.3}&\g{0.1}&\g{0.9}&\g{0.2}&\g{0.5}&\g{1.2}\\
			\footnotesize C$_1$&\g{401.5}&\g{0.1}&\g{0.8}&\g{0.6}&\g{0.3}&\g{0.3}&\g{0.9}&\g{0.5}&\g{1.3}\\
			\footnotesize C$_2$&\g{401.4}&\g{0.2}&\g{1.8}&\g{0.9}&\g{0.2}&\g{1.0}&\g{0.6}&\g{0.2}&\g{0.2}\\
			\hline
			\multicolumn{10}{c}{ (a)  AWGN   } \\
			
						\rowcolor{gray!50}
			& \scriptsize (\emph{i,j})& \scriptsize  (\emph{i,j}+1)&  \scriptsize (\emph{i,j}+2)&  \scriptsize (\emph{i}+1,\emph{j})& \scriptsize (\emph{i}+1,\emph{j}+1)&  \scriptsize (\emph{i}+1,\emph{j}+2) &  \scriptsize (\emph{i}+2,\emph{j}) &  \scriptsize (\emph{i}+2,\emph{j}+1)&  \scriptsize (\emph{i}+2,\emph{j}+2)
			\\\hline\hline
			
			\footnotesize R& \g{359.6}&\g{152.1}&\g{15.1}&\g{154.8}&\g{92.5}&\g{18.9}&\g{18.6}&\g{17.6}&\g{8.5}\\
			\footnotesize G&\g{359.3}&\g{91.4}&\g{1.0}&\g{100.3}&\g{23.9}&\g{1.8}&\g{0.8}&\g{0.4}&\g{5.1}\\
			\footnotesize B&\g{377.4}&\g{150.7}&\g{15.2}&\g{155.5}&\g{89.3}&\g{18.5}&\g{20.6}&\g{17.5}&\g{8.1}\\
			\hline\hline
			\footnotesize Y&\g{654.4}&\g{185.4}&\g{50.8}&\g{196.1}&\g{60.0}&\g{2.9}&\g{49.7}&\g{9.1}&\g{19.0}\\
			\footnotesize C$_1$&\g{274.6}&\g{143.2}&\g{42.5}&\g{144.9}&\g{99.3}&\g{22.1}&\g{48.3}&\g{24.5}&\g{6.4}\\
			\footnotesize C$_2$&\g{167.2}&\g{65.5}&\g{37.6}&\g{69.7}&\g{46.4}&\g{20.0}&\g{41.4}&\g{20.0}&\g{9.1}\\
			\hline
			\multicolumn{10}{c}{ (b)  HA  } \\
			
			\rowcolor{gray!50}
			& \scriptsize (\emph{i,j})& \scriptsize  (\emph{i,j}+1)&  \scriptsize (\emph{i,j}+2)&  \scriptsize (\emph{i}+1,\emph{j})& \scriptsize (\emph{i}+1,\emph{j}+1)&  \scriptsize (\emph{i}+1,\emph{j}+2) &  \scriptsize (\emph{i}+2,\emph{j}) &  \scriptsize (\emph{i}+2,\emph{j}+1)&  \scriptsize (\emph{i}+2,\emph{j}+2)
			\\\hline\hline
			
			\footnotesize R&\g{336.4}&\g{126.8}&\g{19.4}&\g{129.9}&\g{52.9}&\g{21.6}&\g{20.7}&\g{22.4}&\g{18.7}\\
			\footnotesize G&\g{295.5}&\g{92.5}&\g{0.5}&\g{95.6}&\g{20.6}&\g{1.8}&\g{0.7}&\g{1.5}&\g{4.3}\\
			\footnotesize B&\g{350.5}&\g{125.9}&\g{18.1}&\g{130.4}&\g{50.7}&\g{20.8}&\g{20.0}&\g{20.9}&\g{17.5}\\
			\hline\hline
			\footnotesize Y&\g{715.6}&\g{170.9}&\g{32.3}&\g{178.6}&\g{2.6}&\g{5.4}&\g{34.0}&\g{7.1}&\g{20.5}\\
			\footnotesize C$_1$&\g{168.4}&\g{108.3}&\g{41.3}&\g{110.1}&\g{73.4}&\g{28.2}&\g{44.1}&\g{29.4}&\g{9.7}\\
			\footnotesize C$_2$&\g{98.3}&\g{66.0}&\g{27.9}&\g{67.3}&\g{48.1}&\g{21.4}&\g{29.9}&\g{22.4}&\g{10.4}\\
			\hline
			\multicolumn{10}{c}{ (c)  RI  } \\
			
			\rowcolor{gray!50}
			& \scriptsize (\emph{i,j})& \scriptsize  (\emph{i,j}+1)&  \scriptsize (\emph{i,j}+2)&  \scriptsize (\emph{i}+1,\emph{j})& \scriptsize (\emph{i}+1,\emph{j}+1)&  \scriptsize (\emph{i}+1,\emph{j}+2) &  \scriptsize (\emph{i}+2,\emph{j}) &  \scriptsize (\emph{i}+2,\emph{j}+1)&  \scriptsize (\emph{i}+2,\emph{j}+2)
			\\\hline\hline
			
			\footnotesize R&\g{361.4}&\g{128.4}&\g{18.9}&\g{130.5}&\g{46.4}&\g{20.6}&\g{21.6}&\g{21.5}&\g{19.8}\\
			\footnotesize G&\g{298.9}&\g{93.0}&\g{0.5}&\g{95.1}&\g{19.1}&\g{0.9}&\g{1.0}&\g{0.5}&\g{3.8}\\
			\footnotesize B&\g{370.9}&\g{127.8}&\g{19.3}&\g{130.4}&\g{46.0}&\g{20.6}&\g{21.2}&\g{20.3}&\g{19.0}\\
			\hline\hline
			\footnotesize Y&\g{772.2}&\g{177.7}&\g{33.0}&\g{181.3}&\g{9.6}&\g{9.2}&\g{32.6}&\g{10.9}&\g{21.4}\\
			\footnotesize C$_1$&\g{164.8}&\g{107.1}&\g{43.7}&\g{108.8}&\g{72.8}&\g{29.3}&\g{46.1}&\g{30.2}&\g{10.1}\\
			\footnotesize C$_2$&\g{94.3}&\g{64.4}&\g{28.1}&\g{65.8}&\g{48.2}&\g{21.9}&\g{30.3}&\g{23.1}&\g{11.1}\\
			\hline
			\multicolumn{10}{c}{ (d)  MLRI   } \\
			\rowcolor{gray!50}
			& \scriptsize (\emph{i,j})& \scriptsize  (\emph{i,j}+1)&  \scriptsize (\emph{i,j}+2)&  \scriptsize (\emph{i}+1,\emph{j})& \scriptsize (\emph{i}+1,\emph{j}+1)&  \scriptsize (\emph{i}+1,\emph{j}+2) &  \scriptsize (\emph{i}+2,\emph{j}) &  \scriptsize (\emph{i}+2,\emph{j}+1)&  \scriptsize (\emph{i}+2,\emph{j}+2)
			\\\hline\hline
			
			\footnotesize R&\g{359.9}&\g{47.8}&\g{5.0}&\g{51.9}&\g{21.8}&\g{17.8}&\g{5.1}&\g{19.4}&\g{9.2}\\
			\footnotesize G&\g{354.8}&\g{32.6}&\g{4.4}&\g{36.3}&\g{5.8}&\g{8.4}&\g{6.4}&\g{8.8}&\g{0.6}\\
			\footnotesize B&\g{356.0}&\g{49.6}&\g{6.3}&\g{53.7}&\g{23.6}&\g{18.8}&\g{7.3}&\g{19.4}&\g{9.2}\\
			\hline\hline
			\footnotesize Y&\g{972.3}&\g{69.0}&\g{20.8}&\g{76.4}&\g{3.6}&\g{18.6}&\g{28.9}&\g{17.3}&\g{2.2}\\
			\footnotesize C$_1$&\g{55.1}&\g{33.8}&\g{15.3}&\g{36.0}&\g{26.1}&\g{14.6}&\g{19.0}&\g{16.6}&\g{11.8}\\
			\footnotesize C$_2$&\g{43.3}&\g{27.3}&\g{12.3}&\g{29.4}&\g{21.5}&\g{11.7}&\g{16.0}&\g{13.7}&\g{9.4}\\
			\hline
			\multicolumn{10}{c}{ (e)  RCNN   }
			
		\end{tabular}
	\end{center}
\end{table}
\begin{table}[t]
	\caption{Correlation between pixels. The corresponding correlations of $(\mathbf{R},\mathbf{G},\mathbf{B})$ and $(\mathbf{Y},\mathbf{C}_{1},\mathbf{C}_{2})$ 
	between pixels $(i,j)$ and $(i+s,j+t)$, $s,t = 0,1,2$ first for AWGN (a) with standard deviation $\sigma = 20$, then for its demosaiced versions by HA (b), RI (c),  MLRI (d) and  RCNN (e).}
	\label{Table correspondingpixels4level20}
	\begin{center}
		\renewcommand{\arraystretch}{1} \addtolength{\tabcolsep}{-5pt} 
		\footnotesize
		\begin{tabular}{l |cccccccccccc}
			\rowcolor{gray!50}
			
			& \scriptsize (\emph{i,j}) & \scriptsize  (\emph{i,j}+1) &  \scriptsize (\emph{i,j}+2) &  \scriptsize (\emph{i}+1,\emph{j}) & \scriptsize (\emph{i}+1,\emph{j}+1)&  \scriptsize (\emph{i}+1,\emph{j}+2)  &  \scriptsize (\emph{i}+2,\emph{j})  &  \scriptsize (\emph{i}+2,\emph{j}+1) &  \scriptsize (\emph{i}+2,\emph{j}+2)
			\\\hline\hline
			\footnotesize R&\gl{1.0000}&\gl{0.0015}&\gl{0.0010}& \gl{0.0017}&\gl{0.0002}&\gl{0.0018}&\gl{0.0007}&\gl{0.0005}&\gl{0.0021}\\
			\footnotesize G&\gl{1.0000}&\gl{0.0012}&\gl{0.0028}&\gl{0.0004}&\gl{0.0007}&\gl{0.0023}&\gl{0.0025}&\gl{0.0016}&\gl{0.0010}\\
    		\footnotesize B&\gl{1.0000}&\gl{0.0029}&\gl{0.0002}&\gl{0.0013}&\gl{0.0015}&\gl{0.0001}&\gl{0.0047}&\gl{0.0008}&\gl{0.0047}\\
			\hline\hline
			\footnotesize Y&\gl{1.0000}&\gl{0.0028}&\gl{0.0004}&\gl{0.0007}&\gl{0.0002}&\gl{0.0023}&\gl{0.0005}&\gl{0.0012}&\gl{0.0030}\\
			\footnotesize C$_1$&\gl{1.0000}&\gl{0.0003}&\gl{0.0021}&\gl{0.0016}&\gl{0.0007}&\gl{0.0008}&\gl{0.0024}&\gl{0.0011}&\gl{0.0033}\\
			\footnotesize C$_2$&\gl{1.0000}&\gl{0.0005}&\gl{0.0045}&\gl{0.0023}&\gl{0.0005}&\gl{0.0025}&\gl{0.0014}&\gl{0.0005}&\gl{0.0005}\\
			\hline
			\multicolumn{10}{c}{ (a)  AWGN   } \\
			
			\rowcolor{gray!50}
			& \scriptsize (\emph{i,j}) & \scriptsize  (\emph{i,j}+1) &  \scriptsize (\emph{i,j}+2) &  \scriptsize (\emph{i}+1,\emph{j}) & \scriptsize (\emph{i}+1,\emph{j}+1)&  \scriptsize (\emph{i}+1,\emph{j}+2)  &  \scriptsize (\emph{i}+2,\emph{j})  &  \scriptsize (\emph{i}+2,\emph{j}+1) &  \scriptsize (\emph{i}+2,\emph{j}+2)
			\\\hline\hline
			
			\footnotesize R &\gl{1.0000}&\gl{0.4229}&\gl{0.0420}&\gl{0.4307}&\gl{0.2574}&\gl{0.0525}&\gl{0.0518}&\gl{0.0489}&\gl{0.0236}\\
			\footnotesize G&\gl{1.0000}&\gl{0.2543}&\gl{0.0029}&\gl{0.2791}&\gl{0.0666}&\gl{0.0050}&\gl{0.0022}&\gl{0.0010}&\gl{0.0142}\\
			\footnotesize B&\gl{1.0000}&\gl{0.3994}&\gl{0.0403}&\gl{0.4122}&\gl{0.2368}&\gl{0.0490}&\gl{0.0545}&\gl{0.0464}&\gl{0.0215}\\
			\hline\hline
			\footnotesize Y &\gl{1.0000}&\gl{0.2834}&\gl{0.0777}&\gl{0.2997}&\gl{0.0918}&\gl{0.0044}&\gl{0.0760}&\gl{0.0138}&\gl{0.0290}\\
			\footnotesize C$_1$&\gl{1.0000}&\gl{0.5215}&\gl{0.1548}&\gl{0.5278}&\gl{0.3619}&\gl{0.0804}&\gl{0.1759}&\gl{0.0892}&\gl{0.0234}\\
			\footnotesize C$_2$&\gl{1.0000}&\gl{0.3919}&\gl{0.2248}&\gl{0.4166}&\gl{0.2776}&\gl{0.1194}&\gl{0.2477}&\gl{0.1198}&\gl{0.0547}\\
			\hline
			\multicolumn{10}{c}{ (b)  HA  } \\
			
			\rowcolor{gray!50}
			&\scriptsize (\emph{i,j}) &\scriptsize  (\emph{i,j}+1) & \scriptsize (\emph{i,j}+2) &  \scriptsize (\emph{i}+1,\emph{j}) & \scriptsize (\emph{i}+1,\emph{j}+1)&  \scriptsize (\emph{i}+1,\emph{j}+2) & \scriptsize (\emph{i}+2,\emph{j})  &  \scriptsize (\emph{i}+2,\emph{j}+1) &  \scriptsize (\emph{i}+2,\emph{j}+2)
			\\\hline\hline
			
			\footnotesize R &\gl{1.0000}&\gl{0.3744}&\gl{0.0588}&\gl{0.3893}&\gl{0.1536}&\gl{0.0633}&\gl{0.0671}&\gl{0.0626}&\gl{0.0542}\\
			\footnotesize G & \gl{1.0000}&\gl{0.3099}&\gl{0.0044}&\gl{0.3265}&\gl{0.0681}&\gl{0.0063}&\gl{0.0038}&\gl{0.0040}&\gl{0.0163}\\
			\footnotesize B &\gl{1.0000}&\gl{0.3631}&\gl{0.0579}&\gl{0.3715}&\gl{0.1431}&\gl{0.0631}&\gl{0.0612}&\gl{0.0585}&\gl{0.0523}\\
			\hline\hline
			\footnotesize Y &\gl{1.0000}&\gl{0.2382}&\gl{0.0419}&\gl{0.2510}&\gl{0.0003}&\gl{0.0058}&\gl{0.0407}&\gl{0.0129}&\gl{0.0298}\\
			\footnotesize C$_1$ &\gl{1.0000}&\gl{0.6422}&\gl{0.2442}&\gl{0.6548}&\gl{0.4345}&\gl{0.1655}&\gl{0.2639}&\gl{0.1746}&\gl{0.0568}\\
			\footnotesize C$_2$ &\gl{1.0000}&\gl{0.6690}&\gl{0.2795}&\gl{0.6846}&\gl{0.4904}&\gl{0.2188}&\gl{0.3012}&\gl{0.2291}&\gl{0.1075}\\
			\hline
			\multicolumn{10}{c}{ (c)  RI  } \\
			
			\rowcolor{gray!50}
			& \scriptsize (\emph{i,j}) & \scriptsize  (\emph{i,j}+1) &  \scriptsize (\emph{i,j}+2) &  \scriptsize (\emph{i}+1,\emph{j}) & \scriptsize (\emph{i}+1,\emph{j}+1)&  \scriptsize (\emph{i}+1,\emph{j}+2)  &  \scriptsize (\emph{i}+2,\emph{j})  &  \scriptsize (\emph{i}+2,\emph{j}+1) &  \scriptsize (\emph{i}+2,\emph{j}+2)
			\\\hline\hline
			
			\footnotesize R &\gl{1.0000}&\gl{0.3496}&\gl{0.0516}&\gl{0.3624}&\gl{0.1213}&\gl{0.0544}&\gl{0.0632}&\gl{0.0543}&\gl{0.0546}\\
			\footnotesize G &\gl{1.0000}&\gl{0.3077}&\gl{0.0001}&\gl{0.3221}&\gl{0.0623}&\gl{0.0039}&\gl{0.0099}&\gl{0.0019}&\gl{0.0145}\\
			\footnotesize B &\gl{1.0000}&\gl{0.3449}&\gl{0.0567}&\gl{0.3525}&\gl{0.1225}&\gl{0.0589}&\gl{0.0624}&\gl{0.0561}&\gl{0.0567}\\
			\hline\hline
			\footnotesize Y &\gl{1.0000}&\gl{0.2271}&\gl{0.0404}&\gl{0.2371}&\gl{0.0164}&\gl{0.0103}&\gl{0.0366}&\gl{0.0165}&\gl{0.0305}\\
			\footnotesize C$_1$ &\gl{1.0000}&\gl{0.6479}&\gl{0.2625}&\gl{0.6632}&\gl{0.4400}&\gl{0.1748}&\gl{0.2868}&\gl{0.1863}&\gl{0.0632}\\
			\footnotesize C$_2$ &\gl{1.0000}&\gl{0.6806}&\gl{0.2959}&\gl{0.6965}&\gl{0.5121}&\gl{0.2343}&\gl{0.3200}&\gl{0.2472}&\gl{0.1208}\\
			\hline
			\multicolumn{10}{c}{ (d)  MLRI   } \\

			\rowcolor{gray!50}
			& \scriptsize (\emph{i,j}) & \scriptsize  (\emph{i,j}+1) &  \scriptsize (\emph{i,j}+2) &  \scriptsize (\emph{i}+1,\emph{j}) & \scriptsize (\emph{i}+1,\emph{j}+1)&  \scriptsize (\emph{i}+1,\emph{j}+2)  &  \scriptsize (\emph{i}+2,\emph{j})  &  \scriptsize (\emph{i}+2,\emph{j}+1) &  \scriptsize (\emph{i}+2,\emph{j}+2)
			\\\hline\hline
			
			\footnotesize R &\gl{1.0000}&\gl{0.1328}&\gl{0.0138}&\gl{0.1441}&\gl{0.0605}&\gl{0.0493}&\gl{0.0141}&\gl{0.0538}&\gl{0.0256}\\
			\footnotesize G &\gl{1.0000}&\gl{0.0919}&\gl{0.0125}&\gl{0.1022}&\gl{0.0164}&\gl{0.0237}&\gl{0.0181}&\gl{0.0246}&\gl{0.0016}\\
			\footnotesize B&\gl{1.0000}&\gl{0.1393}&\gl{0.0176}&\gl{0.1508}&\gl{0.0662}&\gl{0.0527}&\gl{0.0206}&\gl{0.0546}&\gl{0.0260}\\
			\hline\hline
			\footnotesize Y &\gl{1.0000}&\gl{0.0709}&\gl{0.0214}&\gl{0.0786}&\gl{0.0037}&\gl{0.0192}&\gl{0.0298}&\gl{0.0178}&\gl{0.0022}\\
			\footnotesize C$_1$&\gl{1.0000}&\gl{0.6129}&\gl{0.2773}&\gl{0.6539}&\gl{0.4730}&\gl{0.2649}&\gl{0.3443}&\gl{0.3003}&\gl{0.2143}\\
			\footnotesize C$_2$&\gl{1.0000}&\gl{0.6302}&\gl{0.2851}&\gl{0.6789}&\gl{0.4963}&\gl{0.2697}&\gl{0.3688}&\gl{0.3171}&\gl{0.2161}\\
			\hline
			\multicolumn{10}{c}{ (e)  RCNN   }

		\end{tabular}
	\end{center}
\end{table}

The color distortion caused by denoising in the YC$_1$C$_2$ space is much less than that in the RGB space, and this transformation does not change the properties of independent identically distributed noise. This explains why it is generally used for color image denoising. We further analyze the properties of residual noise in the YC$_1$C$_2$ color space.

\begin{table}[t]
\caption{Correlation between channels. Covariance (each first row) and corresponding correlation (each second row) of the three color channels (R, G, and B) of the demosaicing noise when the initial CFA white noise satisfies $\sigma = 20$. See Figure \ref{Fig: noisyimageskind20} for an illustration.}
\label{Table Correlation3channels4level20}
\begin{center}
\renewcommand{\arraystretch}{0.9} \addtolength{\tabcolsep}{-2pt}
\small
\begin{tabular}{l|rrrr l|rrrr }
&R&G&B&&&R&G&B 
\\\cmidrule(lr){1-4} \cmidrule(lr){6-9}

\multirow{2}*{R}&\g{359.56}&\g{172.02}&\g{93.85}& &\multirow{2}*{R}&\g{336.44}&\g{206.29}&\g{175.01}\\
&\gl{1.0000}&\gl{0.4786}&\gl{0.2548}&& 
&\gl{1.0000}&\gl{0.6542}&\gl{0.5097}\\
\cmidrule(lr){1-4} \cmidrule(lr){6-9} 

\multirow{2}*{G}&\g{172.02}&\g{359.30}&\g{167.60}& &\multirow{2}*{G}&\g{206.29}&\g{295.54}&\g{200.96}\\
&\gl{0.4786}&\gl{1.0000}&\gl{0.4551}&&  &\gl{0.6542}&\gl{1.0000}&\gl{0.6244}\\
\cmidrule(lr){1-4} \cmidrule(lr){6-9} 

\multirow{2}*{B}&\g{93.85}&\g{167.60}&\g{377.44}& &\multirow{2}*{B}&\g{175.01}&\g{200.96}&\g{350.46}\\
&\gl{0.2548}&\gl{0.4551}&\gl{1.0000}&&  &\gl{0.5097}&\gl{0.6244}&\gl{1.0000}\\ 
\cmidrule(lr){1-4} \cmidrule(lr){6-9} 

\multicolumn{9}{c}{  } \\
&Y&C$_1$&C$_2$&&&Y& C$_1$&C$_2$
\\\cmidrule(lr){1-4} \cmidrule(lr){6-9} 
  
\multirow{2}*{Y}&\g{654.41}&\g{5.50}&\g{31.47}&
&\multirow{2}*{Y}&\g{715.65}&\g{3.55}&\g{9.10}\\
&\gl{1.0000}&\gl{0.0130}&\gl{0.0951}&&
&\gl{1.0000}&\gl{0.0102}&\gl{0.0343}\\
\cmidrule(lr){1-4} \cmidrule(lr){6-9} 

\multirow{2}*{C$_1$}&\g{5.50}&\g{274.65}&\g{7.71}&
&\multirow{2}*{C$_1$}&\g{3.55}&\g{168.44}&\g{7.12}\\
&\gl{0.0130}&\gl{1.0000}&\gl{0.0360}&&
&\gl{0.0102}&\gl{1.0000}&\gl{0.0554}\\
\cmidrule(lr){1-4} \cmidrule(lr){6-9} 

\multirow{2}*{C$_2$}&\g{31.47}&\g{7.71}&\g{167.23}&
&\multirow{2}*{C$_2$}&\g{9.10}&\g{7.12}&\g{98.35}\\
&\gl{0.0951}&\gl{0.0360}&\gl{1.0000}&&
&\gl{0.0343}&\gl{0.0554}&\gl{1.0000}\\
\cmidrule(lr){1-4} \cmidrule(lr){6-9}

\multicolumn{4}{c}{(a)  HA  }&&\multicolumn{4}{c}{(b)  RI  }\\

&R&G&B&&&R&G&B 
\\\cmidrule(lr){1-4} \cmidrule(lr){6-9}
\multirow{2}*{R}&\g{361.42}&\g{224.39}&\g{201.41}& &\multirow{2}*{R}&\g{359.90}&\g{320.44}&\g{302.85}\\ 
&\gl{1.0000}&\gl{0.6826}&\gl{0.5501}&& 
&\gl{1.0000}&\gl{0.8967}&\gl{0.8461}\\
\cmidrule(lr){1-4} \cmidrule(lr){6-9}  

\multirow{2}*{G}&\g{224.39}&\g{298.94}&\g{216.86}& &\multirow{2}*{G}&\g{320.44}&\g{354.83}&\g{299.85}\\ 
&\gl{0.6826}&\gl{1.0000}&\gl{0.6512}&&  &\gl{0.8967}&\gl{1.0000}&\gl{0.8437}\\
\cmidrule(lr){1-4} \cmidrule(lr){6-9}

\multirow{2}*{B}&\g{201.41}&\g{216.86}&\g{370.92}& &\multirow{2}*{B}&\g{302.85}&\g{299.85}&\g{355.99}\\ 
&\gl{0.5501}&\gl{0.6512}&\gl{1.0000}&&  &\gl{0.8461}&\gl{0.8437}&\gl{1.0000}\\
\cmidrule(lr){1-4} \cmidrule(lr){6-9}

\multicolumn{9}{c}{  } \\
&Y&C$_1$&C$_2$&&&Y& C$_1$&C$_2$
\\\cmidrule(lr){1-4} \cmidrule(lr){6-9} 

\multirow{2}*{Y}&\g{772.20}&\g{0.80}&\g{22.64}&
&\multirow{2}*{Y}&\g{972.34}&\g{10.00}&\g{1.97}\\
&\gl{1.0000}&\gl{0.0023}&\gl{0.0839}&&
&\gl{1.0000}&\gl{0.0432}&\gl{0.0096}\\
\cmidrule(lr){1-4} \cmidrule(lr){6-9}

\multirow{2}*{C$_1$}&\g{0.80}&\g{164.76}&\g{7.09}&
&\multirow{2}*{C$_1$}&\g{10.00}&\g{55.09}&\g{10.75}\\
&\gl{0.0023}&\gl{1.0000}&\gl{0.0569}&&
&\gl{0.0432}&\gl{1.0000}&\gl{0.2202}\\
\cmidrule(lr){1-4} \cmidrule(lr){6-9}

\multirow{2}*{C$_2$}&\g{22.64}&\g{7.09}&\g{94.33}&
&\multirow{2}*{C$_2$}&\g{1.97}&\g{10.75}&\g{43.29}\\
&\gl{0.0839}&\gl{0.0569}&\gl{1.0000}&&
&\gl{0.0096}&\gl{0.2202}&\gl{1.0000}\\
\cmidrule(lr){1-4} \cmidrule(lr){6-9}

\multicolumn{4}{c}{(c) MLRI}&&\multicolumn{4}{c}{(d) RCNN }

\end{tabular}
\end{center}
	
\end{table}

From Figure~\ref{Fig: noisyimageskind20} one can see that the AWG noise is isotropic whereas the demosaiced noise is not isotropic anymore in the RGB space.
The noise is elongated in the brightness direction $\mathbf{Y}= \frac{\mathbf{R}+\mathbf{G}+\mathbf{B}}{\sqrt3}$, and compressed in other directions. Furthermore,  the noise becomes blurred after demosaicking.  This indicates that the demosaiced noise  is correlated between adjacent pixels. 
This is also verified in Table \ref{Table Correlationpixels4level20} which illustrates the variances and covariances of AWGN and demosaicked noise with $\sigma =20$  both in RGB and YC$_1$C$_2$ spaces. 
One can observe that the  statistical properties of AWG noise remains unchanged  while that of demosaicked noise changes obviously after $(\mathbf{R},\mathbf{G},\mathbf{B}) \rightarrow (\mathbf{Y},\mathbf{C}_{1},\mathbf{C}_{2})$ transformation.
The variance of $\mathbf{Y}$  is a growing  sequence  for the demosaiced noise obtained by increasingly sophisticated demosaicing: 
$654$ for HA, $715$ for RI,  $772$ for MLRI, $972$ for RCNN. 
Hence, the noise standard deviation on $\mathbf{Y}$  has been multiplied by a factor between $1.27$ and $1.56$. 
In contrast, the demosaiced noise is reduced in the $\mathbf{C}_{1}$  and $\mathbf{C}_{2}$ axes, with its variance passing  from $400$ for AWGN to $168$  and $94$ for RI, and  even  down  to $43$ and  $55$ for RCNN. 
Table \ref{Table Correlationpixels4level20} also shows that the covariances between adjacent pixels are  no longer close to $0$ and that the covariances of demosaicked noise  is an almost descending  sequence   by increasingly sophisticated demosaicing.  
In order to further analyze the correlation between adjacent pixel noises, the correlation coefficients of adjacent pixel noises are calculated and  listed in Table \ref{Table correspondingpixels4level20}. 
 The correlation of AWGN is (almost) $0$ due to the independent properties (see Table \ref{Table correspondingpixels4level20} (a)). However, the demosaiced  noise have a strong correlation in $(\mathbf{R},\mathbf{G},\mathbf{B})$ color space. 
 After transformation, the channel correlation of $\mathbf{Y}$ decreases  significantly and the correlation of $\mathbf{C}_{1}$ and $\mathbf{C}_{2}$ increases.

These observations lead to a simple conclusion: As the computational complexity increases, the $\mathbf{Y}$ component of the demosaiced noise gets closer to white. However, the residual noise on $\mathbf{C}_{1}$ and $\mathbf{C}_{2}$ is strongly spatially correlated, it is therefore a low frequency noise, that will require stronger filtering than white noise to be removed. Since image denoising algorithms are guided by the $\mathbf{Y}$ component~\cite{dabov2007color,lebrun2013nonlocal}, we can denoise with methods designed for white noise, but with a noise parameter  adapted to the increased variance of $\mathbf{Y}$.

To understand why the variance of $\mathbf{Y}$ is far larger than the AWGN it comes from, let us study in Table~\ref{Table Correlation3channels4level20} the correlation between the three  channels $(\mathbf{R}, \mathbf{G}, \mathbf{B})$  in the demosaiced noise of HA, RI, MLRI and RCNN. 
We observe a strong $(\mathbf{R},\mathbf{G},\mathbf{B})$ correlation ranging from 0.4 for HA to 0.89 for RCNN, which is caused by the "tendency to grey" of all demosaicing algorithms (see Figures \ref{Fig: noisyimages} and \ref{Fig: noisyimageskind20}). Assuming that the
demosaiced noisy pixel components (denoted 
$\widetilde{\epsilon}_{\mathbf{R}},\widetilde{\epsilon}_{\mathbf{G}},\widetilde{\epsilon}_{\mathbf{B}}$) have a correlation coefficient close to $1$ then we have 
$$\mathbf{Y}=\frac{\widetilde{\epsilon}_{\mathbf{R}}+\widetilde{\epsilon}_{\mathbf{G}}+\widetilde{\epsilon}_{\mathbf{B}}}{\sqrt{3}} \sim \sqrt{3}\, N(0, {{\sigma}}).$$
This  factor of about $1.7$ corresponds to the case  with maximum correlation. The empirical observation of an optimal factor near $1.5$ responds to a lower correlation between the colors.

All in all, the analysis of the statistical properties of demosaicked noise explains why the $\B\&\A$ scheme with an appropriate parameter $\sigma_{2}=1.5\sigma$ performs similarly to the optimal CMA-ES, and is much better than $\A\&\B$.

\section{Experimental evaluation}
\label{sec:experiment}

To evaluate the proposed framework for denoising and demosaicing, we conducted experiments on simulated images and real images separately. The most classic Imax~\cite{zhang2011color} and Kodak~\cite{franzen1999kodak} datasets were selected for the simulated images. To verify the effect on real raw images, we  also evaluated it on the SIDD dataset~\cite{Abdelhamed2018} and on the DND~\cite{2017DND} benchmark. The former comes with ground truth acquisitions, while the latter allows to evaluate the results by submitting them to the benchmark website.

\subsection{Evaluation of \texorpdfstring{$\A\&\B$}{} and \texorpdfstring{$\B\&1.5\A$}{} strategies on simulated images}
\label{sec: simulation}

  All Imax and Kodak images were corrupted by AWGN with standard deviations ${{\sigma}} = 5, 10, 20, 40, 50,60$.
 
  We compared nine different pipelines, namely:
\begin{itemize}
    \item Best performing $\A\&\B$ and $\B\&1.5\A$ pipelines built by RCNN~\cite{tan2017color} and cfaBM3D or CBM3D~\cite{dabov2007color}.
    
    \item Low cost $\A\&\B$, $\B\&1.5\A$ and CMA-ES pipelines built by MLRI~\cite{kiku2014minimized} and cfaBM3D or CBM3D \cite{dabov2007color}.
   
    \item The CFA denoising framework proposed by Park \emph{et al.} in~\cite{park2009case}, which effectively compresses the signal energy by using a color representation obtained by principal component analysis of the Kodak dataset, and then removes the noise in each channel by BM3D. We combined this framework with RCNN~\cite{tan2017color}.
    
    \item The PCA-CFA filter proposed in~\cite{zhang2009pca} uses principal component analysis (PCA) and spatial and spectral correlation of images to preserve color edges and details. We combined it with DLMM demosaicing \cite{zhang2005color} and RCNN demosaicing \cite{tan2017color}.
    
    \item Since 2016, solving joint demosaicing denoising has typically used deep learning. As a reference, we included JCNN \cite{gharbi2016deep,ehretDEMOSAICK_IPOL2019}, which is one of the classical deep learning algorithms for this problem, for comparison. It is important to emphasize that it was trained on noise standard deviations $\sigma\le20$ only.

\end{itemize}

\begin{table} [t]
	\caption{
	The results of different combinations of denoising and demosaicing methods for the {\bf Imax} image dataset. The best result for each row is \textcolor{red}{red}, the second best result is \textcolor{brown}{brown}, and the third best result is \textcolor{blue}{blue}.
	}
	\label{Table ImaxCompare}
	\begin{center}
		\renewcommand{\arraystretch}{1.1} \addtolength{\tabcolsep}{-4pt} 
		\footnotesize
		\rowcolors{2}{gray!25}{white}
		\begin{tabular}{l| c c c c c |c c | c |c}
			\rowcolor{gray!50}
			\hline 
			& \multicolumn{5}{c|}{$\A\&\B$} & \multicolumn{2}{c|}{$\B\&1.5\A$} & CMA-ES & \\ 
			\hline
			\rowcolor{gray!50}
		    $\sigma$ & cfaBM3D+ & cfaBM3D+ & Park+ & PCA+ & PCA+ & RCNN+ & MLRI+ & cfaBM3D+ & JCNN\\
			\rowcolor{gray!50}
			& MLRI~ & RCNN~ & RCNN & DLMM & RCNN & CBM3D & CBM3D & MLRI+ &\\
						\rowcolor{gray!50}
			&  & &  &  & &  &  & CBM3D &\\\hline
			$ 5$ & 34.20 & \textcolor{brown}{35.21} & 32.86 & 32.69 & \textcolor{blue}{34.87} & \textcolor{red}{35.44} & 34.64 & 34.66 & 33.48 \\
			
			$10$ & 31.68 & 32.26 & 30.06 & 30.73 & 31.89 & \textcolor{brown}{32.77} & 32.35 & \textcolor{blue}{32.43} & \textcolor{red}{33.09} \\
			
			$20$ & 28.48 & 28.73 & 26.86 & 27.57 & 27.99 & \textcolor{brown}{29.54} & 29.30 & \textcolor{blue}{29.36} & \textcolor{red}{29.79}\\
			
			$40$ & 24.90 & 24.92 & 23.86 & 23.50 & 23.57 & \textcolor{brown}{25.69} & \textcolor{blue}{25.46} & \textcolor{red}{25.74} & --\\
			
			$50$ & 23.62 & 23.59 & 22.67 & 22.08 & 22.10 & \textcolor{brown}{24.27} & \textcolor{blue}{24.01} & \textcolor{red}{24.36} & --\\
			
			$60$ & 22.49 & 22.43 & 21.75 & 20.89 & 20.89 & \textcolor{brown}{23.02} &  \textcolor{blue}{22.76} & \textcolor{red}{23.16} & --\\

			\hline
    	    Av & 27.56 & 27.86 & 26.34 & 26.24 & 26.89 & \textcolor{red}{28.46} & \textcolor{blue}{28.09} & \textcolor{brown}{28.29} & --\\
			\hline
		\end{tabular}
	\end{center}
\end{table}
\begin{table}[t]
	\caption{
	The results of different combinations of denoising and demosaicing methods for the {\bf Kodak} image dataset. The best result for each row is \textcolor{red}{red}, the second best result is \textcolor{brown}{brown}, and the third best result is \textcolor{blue}{blue}.
	}
	\label{Table KodakCompare}
	\begin{center}
		\renewcommand{\arraystretch}{1.1} \addtolength{\tabcolsep}{-4pt} \footnotesize
		\rowcolors{2}{gray!25}{white}
		\begin{tabular}{l|c c c c c |c c |c| c}
			\rowcolor{gray!50}
			\hline 
			& \multicolumn{5}{c|}{$\A\&\B$} & \multicolumn{2}{c|}{$\B\&1.5\A$} & CMA-ES & \\ 
			\hline
			\rowcolor{gray!50}
		    $\sigma$ & cfaBM3D+ & cfaBM3D+ & Park+ & PCA+ & PCA+ & RCNN+ & MLRI+ & cfaBM3D+ & JCNN\\
			\rowcolor{gray!50}
			& MLRI~ & RCNN~ & RCNN & DLMM & RCNN & CBM3D & CBM3D & MLRI+ &\\
						\rowcolor{gray!50}
			&  & &  &  & &  &  & CBM3D &\\\hline
			
			$ 5$ & 35.08 & \textcolor{brown}{36.10} & 34.87 & 34.99 & 35.42 & \textcolor{red}{36.58}& 35.77 & \textcolor{blue}{35.78} & 34.13\\
			
			$10$ & 32.15 & 32.56 & 30.85 & 31.83 & 32.01 & \textcolor{red}{33.36} & 32.99 & \textcolor{blue}{33.02} & \textcolor{brown}{33.27}\\
			
			$20$ & 28.91 & 29.03 & 27.42 & 28.11 & 28.14 & \textcolor{red}{30.12} & 29.85 & \textcolor{blue}{29.91}  &\textcolor{brown}{29.95} \\
			
			$40$ & 25.84 & 25.85 & 24.88 & 24.15 & 24.08 & \textcolor{red}{26.82} & \textcolor{blue}{26.53} & \textcolor{brown}{26.72} & --\\
			
			$50$ & 24.83 & 24.83 & 23.91 & 22.85 & 22.77 & \textcolor{red}{25.67} & \textcolor{blue}{25.33} & \textcolor{brown}{25.61} & --\\
			
			$60$ & 23.90 & 23.89 & 23.19 & 21.77 & 21.70 & \textcolor{red}{24.62} & \textcolor{blue}{24.26} & \textcolor{brown}{24.60} & --\\
			
			\hline
			Av & 28.45 & 28.71 & 27.52 & 27.28 & 27.35 & \textcolor{red}{29.53} & \textcolor{blue}{29.12} & \textcolor{brown}{29.27} & --\\
			\hline
		\end{tabular}
	\end{center}
\end{table}
\begin{figure*}[t]
	\footnotesize
	\centering
	\addtolength{\tabcolsep}{-5pt} 
	\begin{tabular}{cccc}
		\includegraphics[width=.23\linewidth,trim={0 0 140 0},clip]{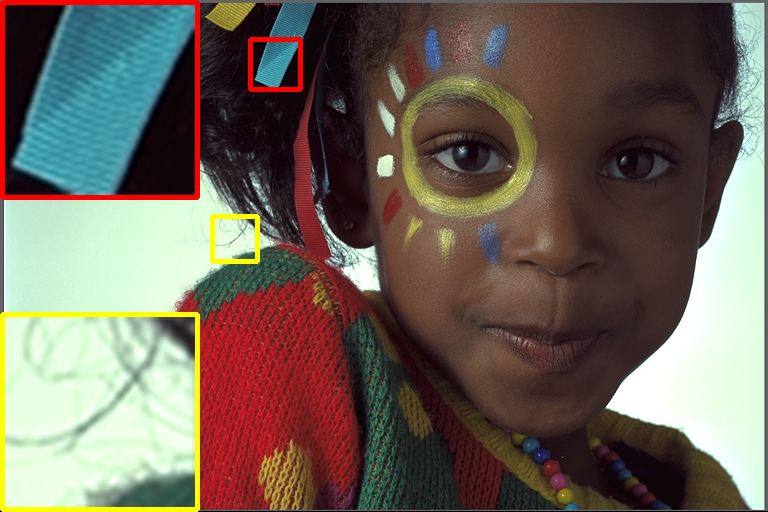}
		& 
		\overimgxx[width=.23\linewidth,trim={0 0 140 0},clip]{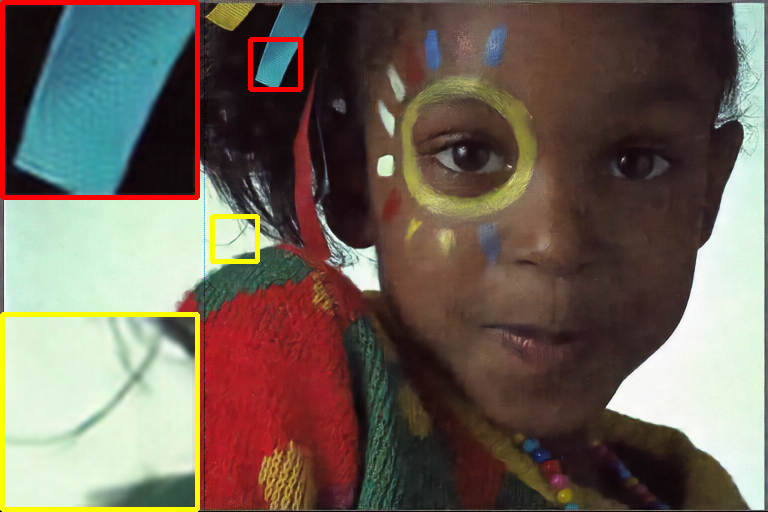}{30.53dB}
	    &
		\overimgxx[width=.23\linewidth,trim={0 0 140 0},clip]{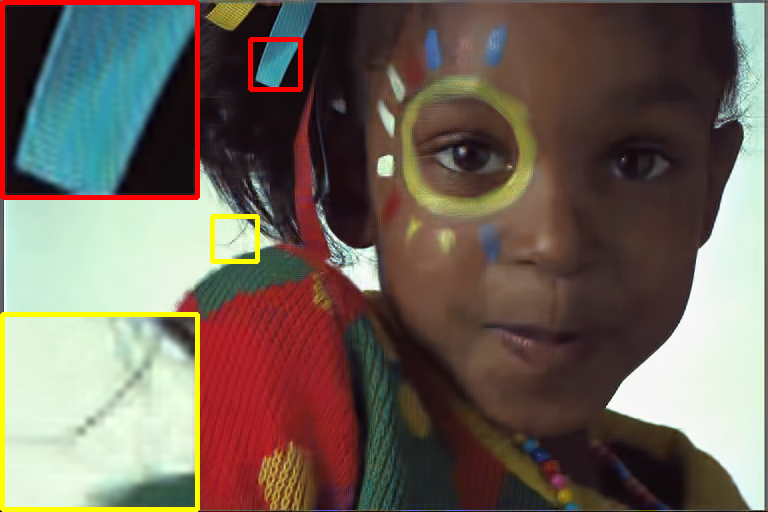}{30.39dB}
		& 
		\overimgxx[width=.23\linewidth,trim={0 0 140 0},clip]{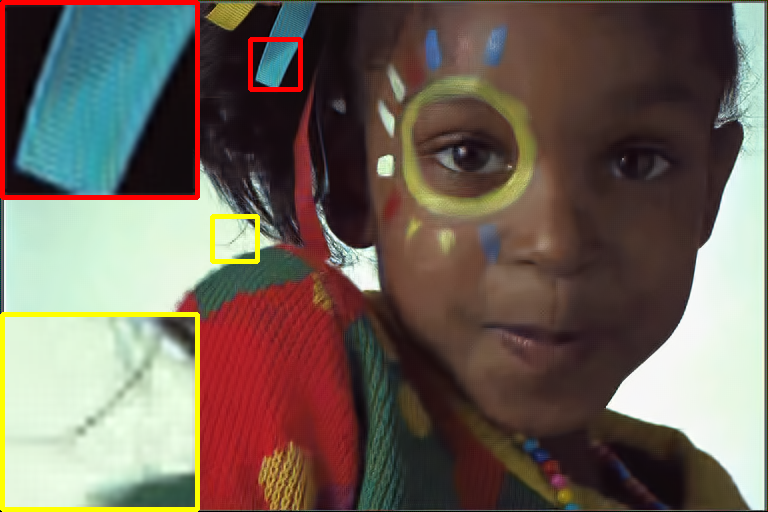}{30.47dB}
		\\
		Ground Truth & JCNN~\cite{gharbi2016deep}  & cfaBM3D+MLRI & cfaBM3D+RCNN  \\
		& & ($\A\&\B$) & ($\A\&\B$) \\
		
		& 
		\overimgxx[width=.23\linewidth,trim={0 0 140 0},clip]{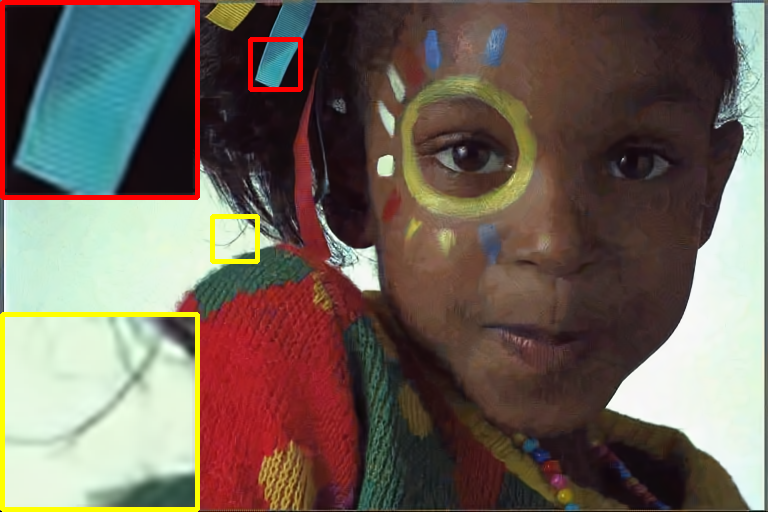}{30.92dB}
		&
		\overimgxx[width=.23\linewidth,trim={0 0 140 0},clip]{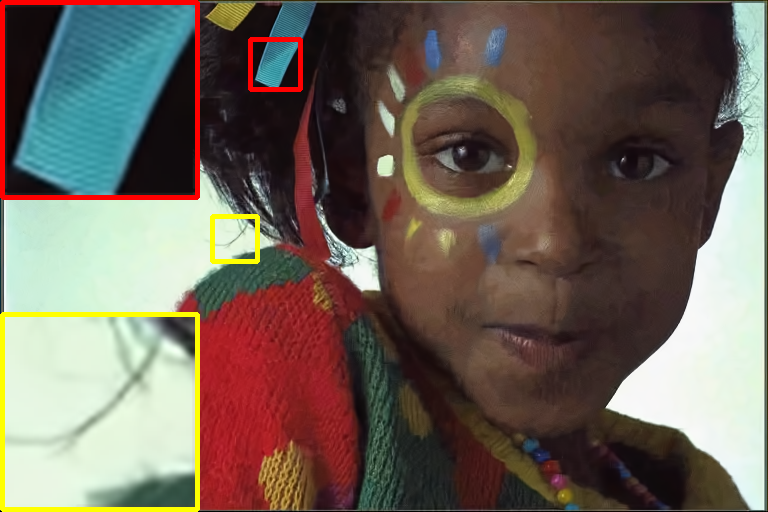}{31.03dB}
		&
		\overimgxx[width=.23\linewidth,trim={0 0 140 0},clip]{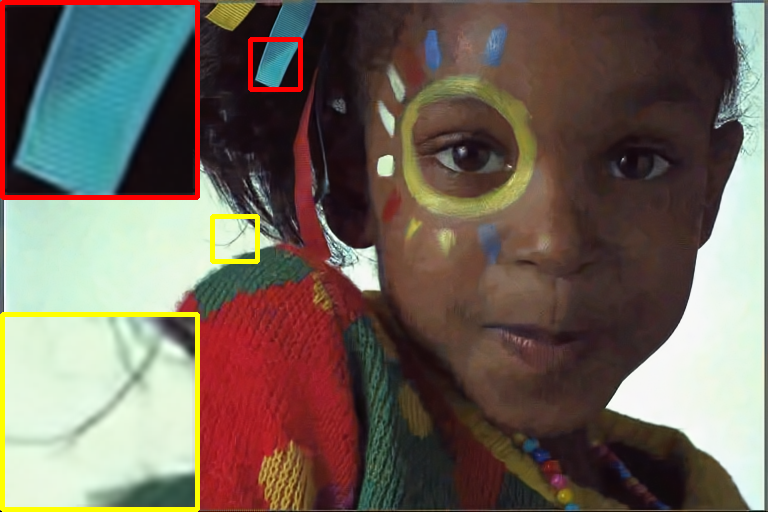}{30.96dB}\\ 
		
		 & MLRI+CBM3D  & RCNN+CBM3D &  MLRI+CBM3D  \\
		 & ($\B\&1.5\A$) & ($\B\&1.5\A$) & (CMA-ES) \\
	\end{tabular}

	\caption{Demosaicing and denoising results on an image from the Kodak dataset with $\sigma=20$. We compare the two schemes of $\A\&\B$, cfaBM3D+MLRI and cfaBM3D+RCNN, the two schemes of $\B\&1.5\A$, MLRI+CBM3D and RCNN+CBM3D, and the MLRI+CBM3D schemes optimized by the CMA-ES algorithm. As a reference we also include the result of JCNN, a joint CNN method.
	}
	\label{fig:compare_kodak_01}
	
\end{figure*}

Table \ref{Table ImaxCompare} shows that RCNN+1.5CBM3D obtains the optimum on average.
It comes to no surprise that JCNN \cite{gharbi2016deep,ehretDEMOSAICK_IPOL2019} performs slightly better than the other methods on the Imax dataset. 
Table \ref{Table KodakCompare} shows that the $\B\&1.5\A$ method RCNN + 1.5CBM3D yields the best results on the Kodak dataset. And when the noise increases, the 'low-cost' MLRI+1.5CBM3D also achieves impressive results. 
However, it is restricted to a limited  range of noise levels and cannot handle the noise levels outside  the training range. Furthermore,  it requires much more memory and computation.
In summary,  $\B\&1.5\A$ methods are more robust and have a better performance than cfaBM3D+RCNN. 
All $\B\&1.5\A$ methods outperform the $\A\&\B$ methods Park+RCNN \cite{park2009case}, PCA+DLMM \cite{zhang2009pca} and PCA+RCNN \cite{zhang2009pca}.

\begin{figure*}[t]
	\footnotesize
	\centering
	\addtolength{\tabcolsep}{-5pt} 
	\begin{tabular}{cccc}
		\includegraphics[width=.23\linewidth,trim={0 0 140 0},clip]{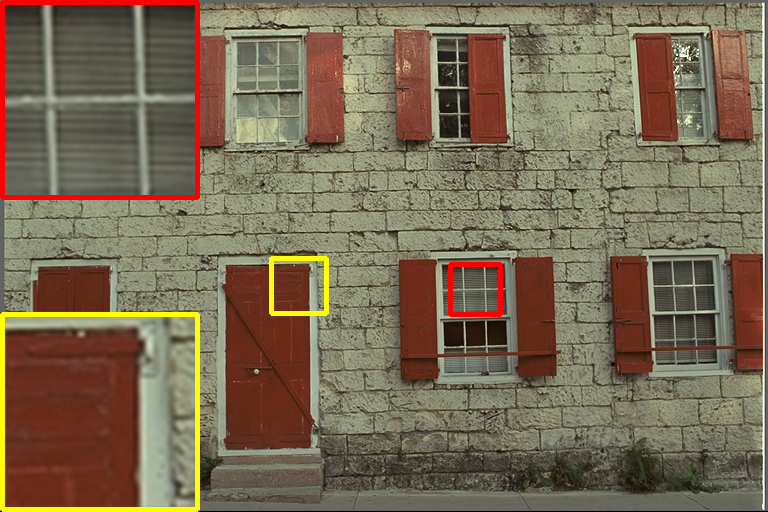}
		& 
		\overimgxx[width=.23\linewidth,trim={0 0 140 0},clip]{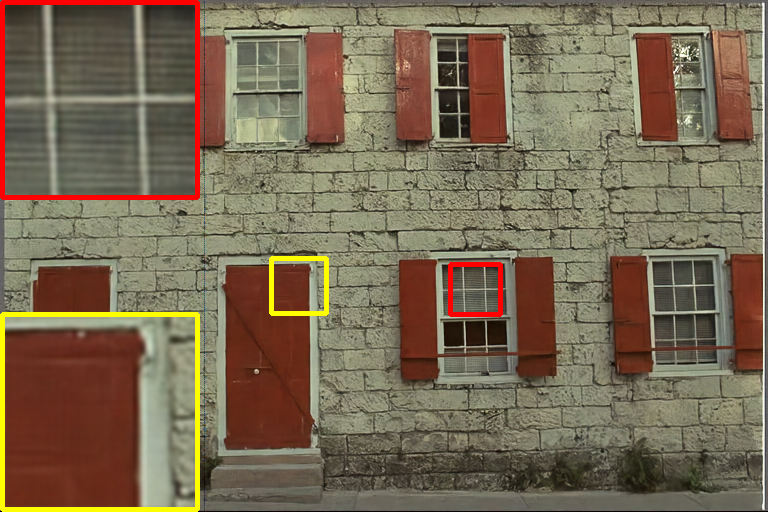}{31.01dB} 
	    &
		\overimgxx[width=.23\linewidth,trim={0 0 140 0},clip]{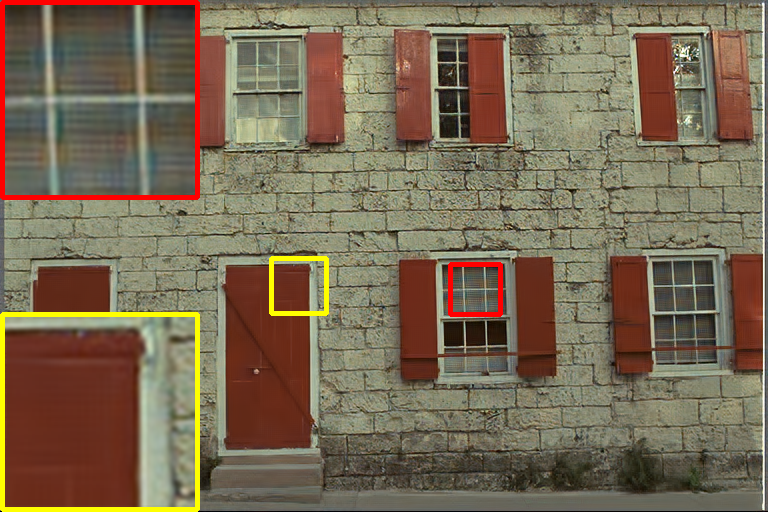}{29.92dB}
		& 
		\overimgxx[width=.23\linewidth,trim={0 0 140 0},clip]{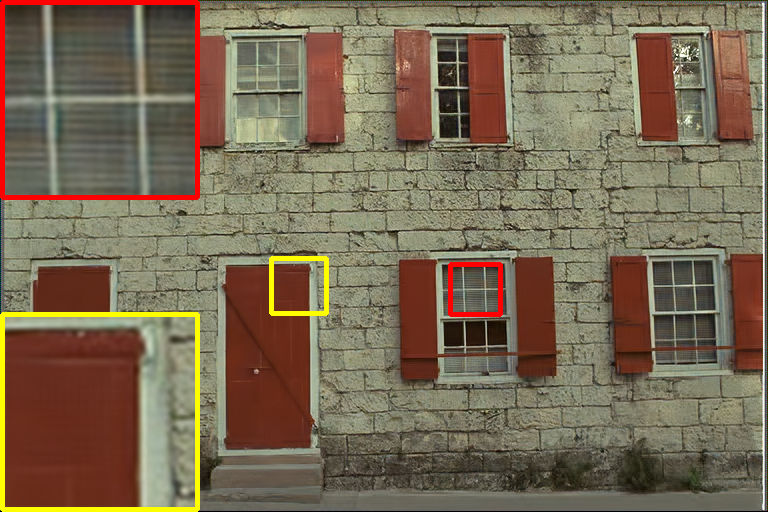}{30.60dB}
		\\
		Ground Truth & JCNN~\cite{gharbi2016deep} &	cfaBM3D+MLRI & cfaBM3D+RCNN \\
		& &($\A\&\B$)& ($\A\&\B$) \\
		
		& 
		\overimgxx[width=.23\linewidth,trim={0 0 140 0},clip]{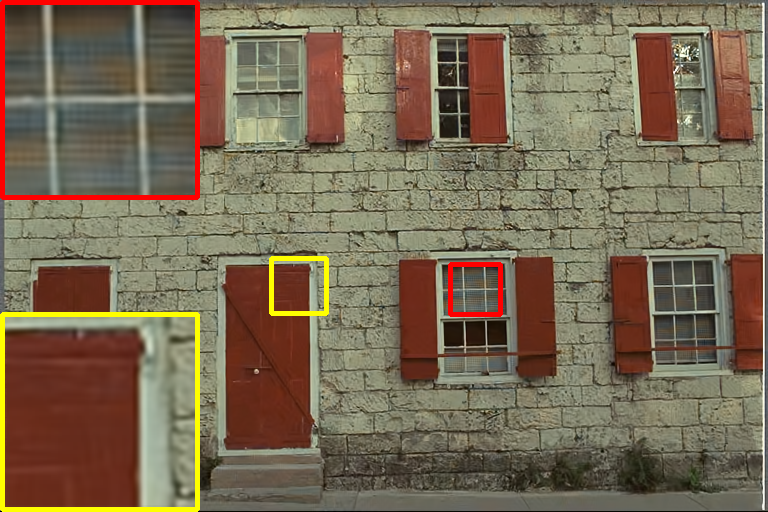}{30.69dB}
		&
		\overimgxx[width=.23\linewidth,trim={0 0 140 0},clip]{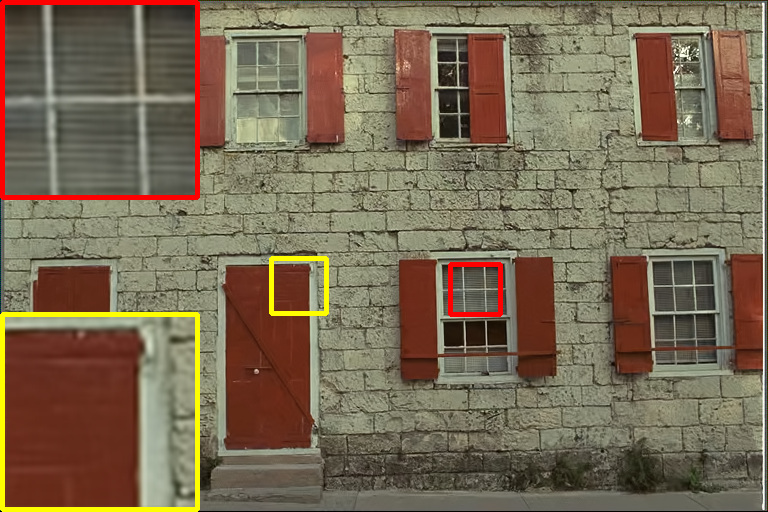}{31.23dB}
		&
		\overimgxx[width=.23\linewidth,trim={0 0 140 0},clip]{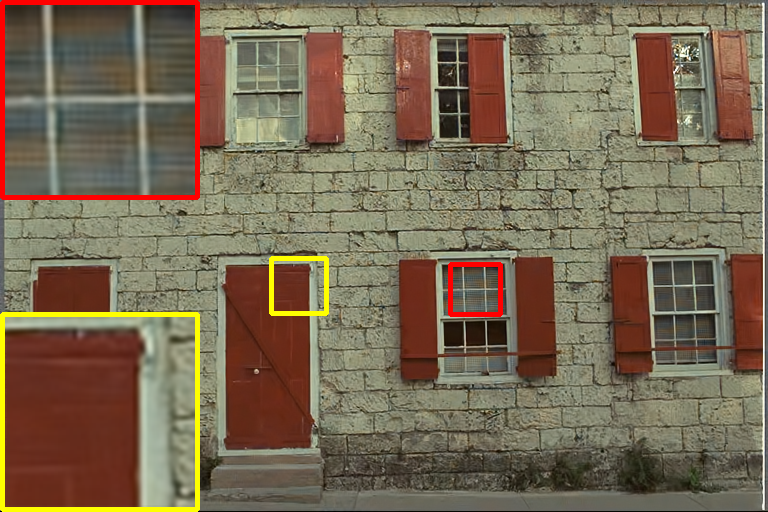}{30.72dB}\\ 
		
		 & MLRI+CBM3D & RCNN+CBM3D &  MLRI+CBM3D \\
		 & ($\B\&1.5\A$) &  ($\B\&1.5\A$) & (CMA-ES) \\
	\end{tabular}
	\caption{Demosaicing and denoising results on an image from the Kodak dataset with $\sigma=10$. We compare the two schemes of $\A\&\B$, cfaBM3D+MLRI and cfaBM3D+RCNN, the two schemes of $\B\&1.5\A$, MLRI+CBM3D and RCNN+CBM3D, and the MLRI+CBM3D schemes optimized by the CMA-ES algorithm. As a reference we also include the result of JCNN, a joint CNN method.
	}
	\vspace{-.5em}
	\label{fig:compare_kodak_02} 
	
\end{figure*}
\begin{figure*}[!htpb]
	\footnotesize
	\centering
	\addtolength{\tabcolsep}{-5pt} 
	\begin{tabular}{cccc}
		\includegraphics[width=.23\linewidth,trim={0 0 0 0},clip]{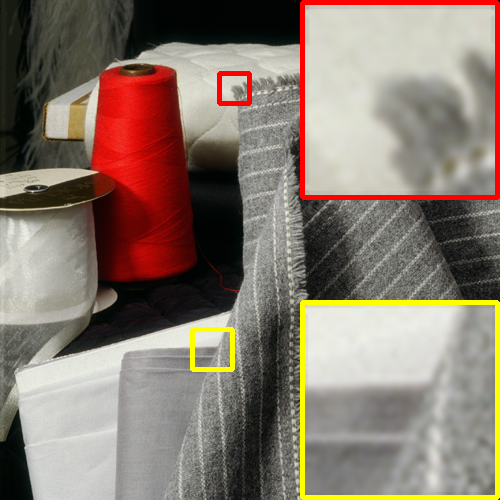}
		& 
		\overimgx[width=.23\linewidth,trim={0 0 0 0},clip]{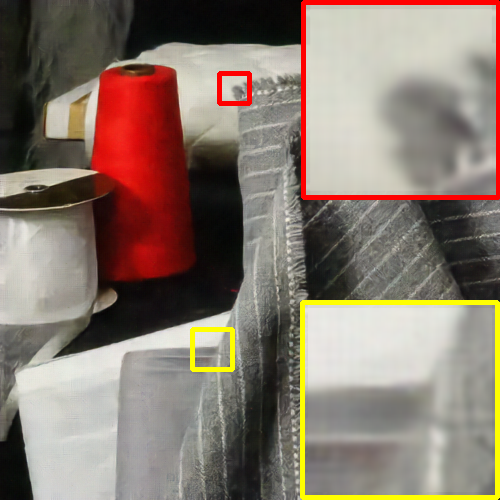}{29.08dB} 
		&
		\overimgx[width=.23\linewidth,trim={0 0 0 0},clip]{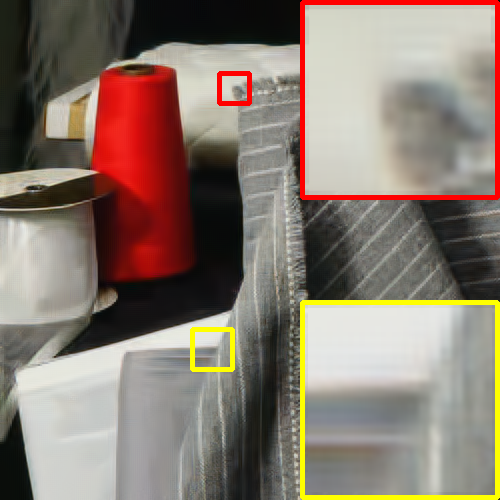}{28.26dB}
		&
		\overimgx[width=.23\linewidth,trim={0 0 0 0},clip]{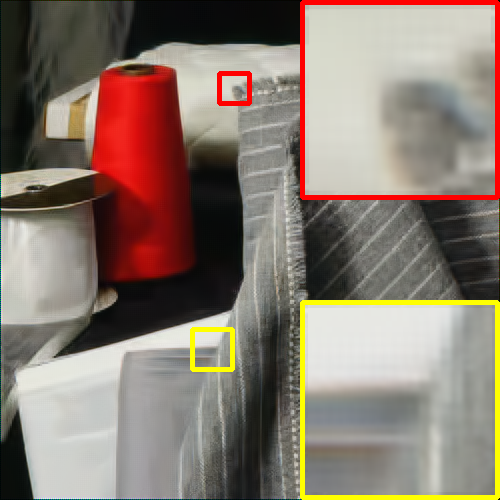}{28.34dB}
		\\
		Ground Truth & 	JCNN~\cite{gharbi2016deep} & cfaBM3D+MLRI  & cfaBM3D+RCNN \\
		&& ($\A\&\B$) & ($\A\&\B$) \\
		
		&
		\overimgx[width=.23\linewidth,trim={0 0 0 0},clip]{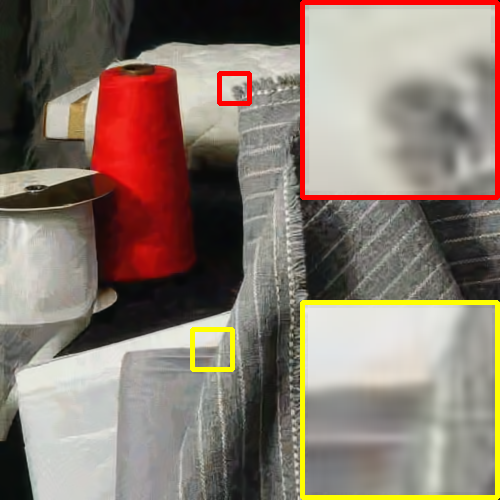}{28.86dB}
		&
		\overimgx[width=.23\linewidth,trim={0 0 0 0},clip]{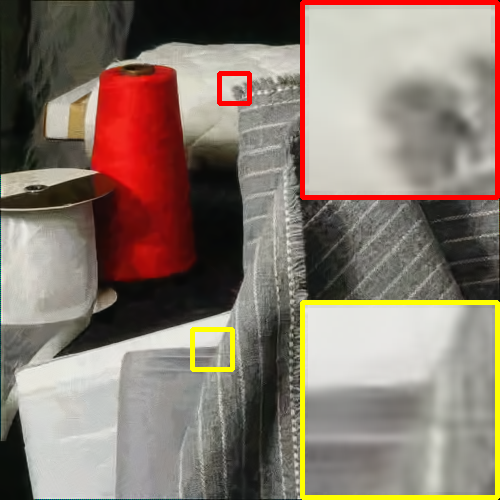}{29.19dB}
		&
		\overimgx[width=.23\linewidth,trim={0 0 0 0},clip]{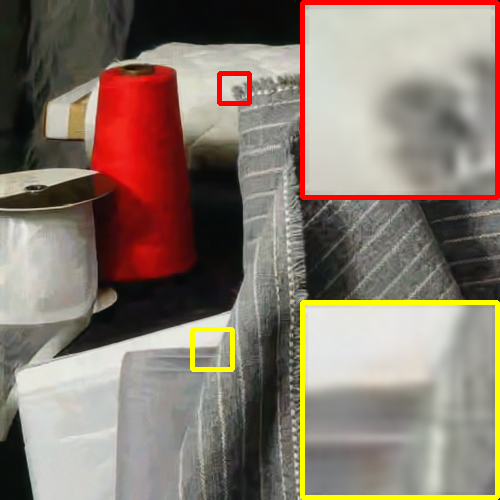}{28.97dB}\\ 
		
		& MLRI+CBM3D & RCNN+CBM3D & MLRI+CBM3D\\
		& ($\B\&1.5\A$) &  ($\B\&1.5\A$) & (CMA-ES) \\
	\end{tabular}

	\caption{Demosaicing and denoising results on an image from the Imax dataset with $\sigma=20$. We compare the two schemes of $\A\&\B$, cfaBM3D+MLRI and cfaBM3D+RCNN, the two schemes of $\B\&1.5\A$, MLRI+CBM3D and RCNN+CBM3D, and the MLRI+CBM3D schemes optimized by the CMA-ES algorithm. As a reference we also include the result of JCNN, a joint CNN method.
	}
	\vspace{-.5em}
	\label{fig:compare_Imax}
\end{figure*}
\begin{figure*}[!htpb]
	\footnotesize
	\centering
	\addtolength{\tabcolsep}{-5pt}
	\begin{tabular}{cccc}
		\includegraphics[width=.23\linewidth,trim={0 0 0 0},clip]{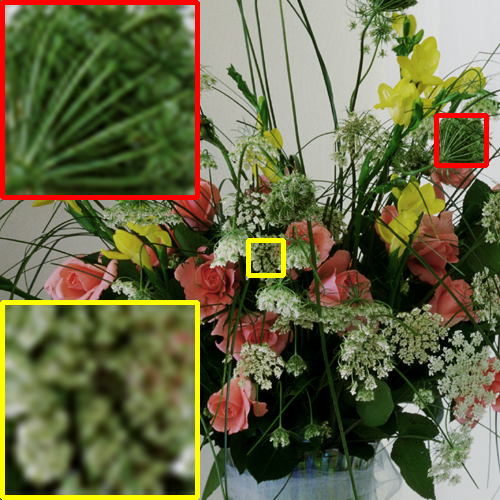}
		& 
		\overimgxx[width=.23\linewidth,trim={0 0 0 0},clip]{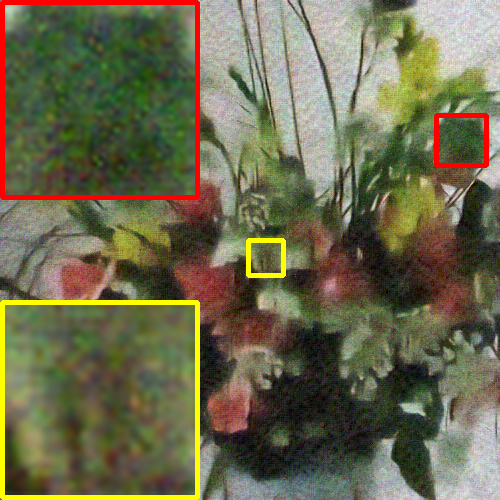}{18.97dB} 
		&
		\overimgxx[width=.23\linewidth,trim={0 0 0 0},clip]{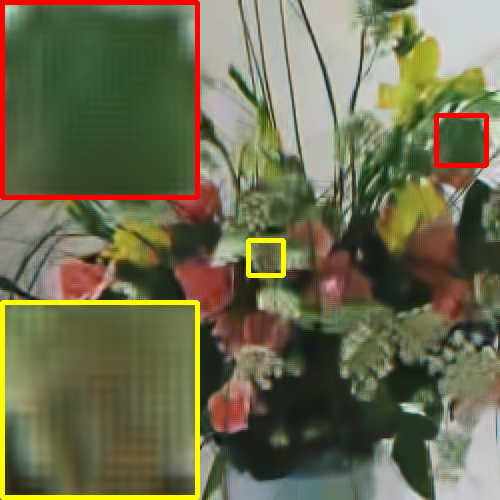}{20.10dB}
		&
		\overimgxx[width=.23\linewidth,trim={0 0 0 0},clip]{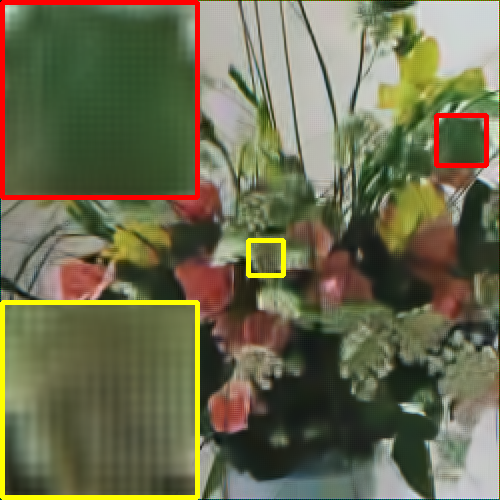}{20.10dB}
		\\ 
		Ground Truth & 	JCNN~\cite{gharbi2016deep} & cfaBM3D+MLRI  & cfaBM3D+RCNN \\
		&&  ($\A\&\B$) & ($\A\&\B$) \\
		
		&
		\overimgxx[width=.23\linewidth,trim={0 0 0 0},clip]{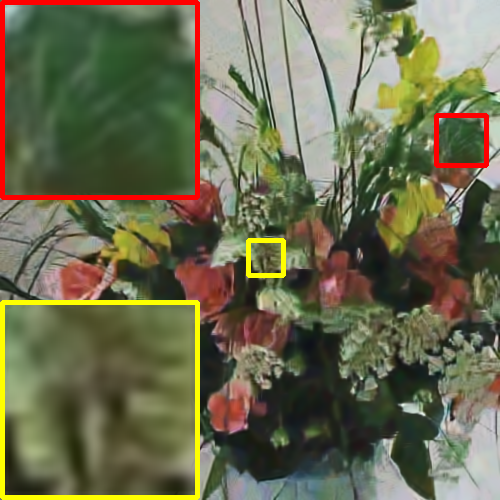}{21.25dB}
		&
		\overimgxx[width=.23\linewidth,trim={0 0 0 0},clip]{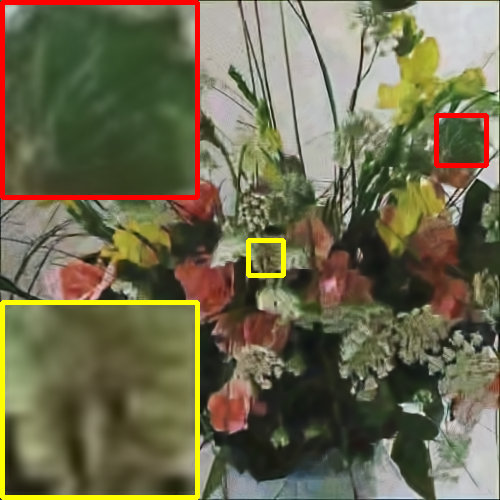}{21.40dB}
		&
		\overimgxx[width=.23\linewidth,trim={0 0 0 0},clip]{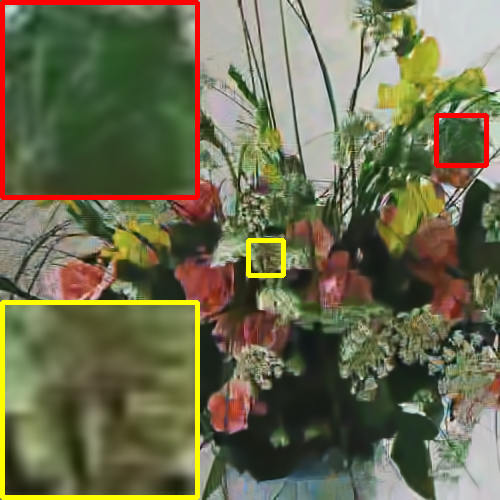}{21.39dB}\\ 
		
		& MLRI+CBM3D & RCNN+CBM3D & MLRI+CBM3D \\
		& ($\B\&1.5\A$) & ($\B\&1.5\A$) & (CMA-ES) \\
	\end{tabular}

	\caption{Demosaicing and denoising results on an image from the Imax dataset with $\sigma=60$. We compare the two schemes of $\A\&\B$, cfaBM3D+MLRI and cfaBM3D+RCNN, the two schemes of $\B\&1.5\A$, MLRI+CBM3D and RCNN+CBM3D, and the MLRI+CBM3D schemes optimized by the CMA-ES algorithm. As a reference we also include the result of JCNN, a joint CNN method.
	} 
	\label{fig:compare_Imax_60}
\end{figure*}

We now examine the visual quality of restored images.
Figures~\ref{fig:compare_kodak_01}-\ref{fig:compare_Imax} compare the visual quality obtained by the main discussed  methods.  
Key parts of images were zoomed-in for a better view.
From the upper-left extract of Figure~\ref{fig:compare_kodak_01}, we can see that textures are well restored by RCNN+1.5CBM3D and MLRI+1.5CBM3D, while they are blurred the cfaBM3D+RCNN and  destroyed by JCNN. 
In the lower-left extract, the girl\rq{}s hairs are oversmoothed by  cfaBM3D+RCNN and JCNN but are well preserved by our proposed method.
In the upper-left and lower-left corner of Figure~\ref{fig:compare_kodak_02}, cfaBM3D+RCNN oversmooths the details and JCNN introduces some artifacts at the window and oversmooths the door. Instead, RCNN+1.5CBM3D preserves the details and does not introduce artifacts. 
The zoomed-in parts of Figure~\ref{fig:compare_Imax} show that JCNN and cfaBM3D+RCNN introduce checkerboard artifacts while methods based on the  $\B\&1.5\A$ scheme do not. 
The advantage of our proposed approach becomes more obvious when dealing with high noise.
There are severe checkerboard artifacts in the images restored by cfaBM3D+MLRI and cfaBM3D+RCNN (see in the bottom left-hand corner of the image of  Figure \ref{fig:compare_Imax_60}),  and the details are oversmoothed (see in the upper left corner  of the image of  Figure \ref{fig:compare_Imax_60}), while our proposed approach not only avoids checkerboard artifacts, but also retains the details. The image restored with JCNN  is very noisy  because JCNN was not trained beyond $\sigma=20$.

As a rule of thumb, the $\B\&\A$   scheme with an appropriate parameter (namely $\B\&1.5 \A$) outperforms the competition in terms of visual quality. This is due to the fact that it efficiently uses spatial and spectral image characteristics to remove noise, preserve edges and fine detail. Indeed, contrary to the $\A\&\B$ schemes, $\B\&1.5 \A$ does  not reduce the resolution of the noisy image. Using a $\A\&\B$ scheme ends up over-smoothing the result.
A comparison of  CPSNRs  and visual quality on these simulated examples leads to conclude that the $\B\&1.5\A$ scheme  is indeed much more robust and better performing than the $\A\&\B$ scheme.

\begin{table}[t]
	\centering
	\renewcommand{\arraystretch}{1.1} \addtolength{\tabcolsep}{-1pt}
	\footnotesize
	\begin{tabular}{c|cccccc}
		Camera & $\sigma$ range & JCNN  & cfaBM3D+ & cfaBM3D+ & MLRI+  &  RCNN+ \\      
               &                &       & MLRI     & RCNN   & CBM3D  &  CBM3D \\
		\hline 
		IP7 & $[5.29, 10.65]$ & 36.79 & 37.30 &  37.43 &  \textcolor{brown}{37.72} & \textcolor{red}{38.37} \\
	    S6 & $[3.71, 38.12]$  & 32.89 & 33.15 &  33.31 &  \textcolor{brown}{33.96} & \textcolor{red}{33.97} \\
		GP & $[3.28, 35.90]$  & 36.42 & 36.78 &  37.15 &  \textcolor{brown}{37.52} & \textcolor{red}{37.58} \\
		N6 & $[4.03, 31.15]$  & 33.38 & 33.96 &  34.16 &  \textcolor{brown}{34.36} & \textcolor{red}{34.21} \\  
		G4 & $[4.66, 13.85]$  & 37.03 & 37.00 &  37.20 &  \textcolor{brown}{37.94} & \textcolor{red}{37.97} \\
		
		\hline
		Av. & $[3.28, 38.12]$ & 35.41 & 35.80 &  36.00 & \textcolor{brown}{36.41} & \textcolor{red}{36.63} \\
	\end{tabular}
	\caption{Average CPSNR results on the SIDD dataset. Note that for each camera, images with different noise levels are being considered. The noise range is $\sigma \in [3.28, 38.12]$. The proposed $\B\&1.5\A$ schemes outperforms the $\A\&\B$ ones.  The best result is in \textcolor{red}{red}, the second best one is in \textcolor{brown}{brown}.}
	\label{tab:real images}
\end{table}

\subsection{Evaluation of \texorpdfstring{$\A\&\B$}{} and \texorpdfstring{$\B\&1.5\A$}{} strategies  on real image datasets}
 
In order to prove the advantage of a $\B\&1.5\A$ strategy on real images, we evaluated its application to the real sRGB images taken from the SIDD dataset~\cite{Abdelhamed2018}. 
In this dataset, the noisy sRGB images and their corresponding ground truth images were acquired by five different  mobile phone models.
We  considered the five most effective demosaicing and denoising schemes among those considered above, namely cfaBM3D+MLRI, cfaBM3D+RCNN, MLRI+1.5CBM3D, RCNN+1.5CBM3D and JCNN.
The noise level was estimated by using the method \cite{Chen2015_noise_level} and provided to the denoising algorithms and JCNN. 
Since the sRGB images used in this experiment are already tone-mapped we   assumed that the resulting noise is approximately homoscedastic. This allowed us to estimate a single noise level per image instead of a noise curve. Thus, a different noise level was computed for each image in the \textbf{SIDD sRGB image dataset}. The noise estimated for all the images  is in the range $\sigma \in [3.28, 38.12]$, and the noise level of most of the images ($\geq 93.75\%$) is no higher than $20$. This  justifies the choice of $\B\&1.5\A$.

\begin{figure*}
	\footnotesize
	\centering
	\addtolength{\tabcolsep}{-3pt}
	\begin{tabular}{ccc}
	\overimgx[width=0.23\linewidth]{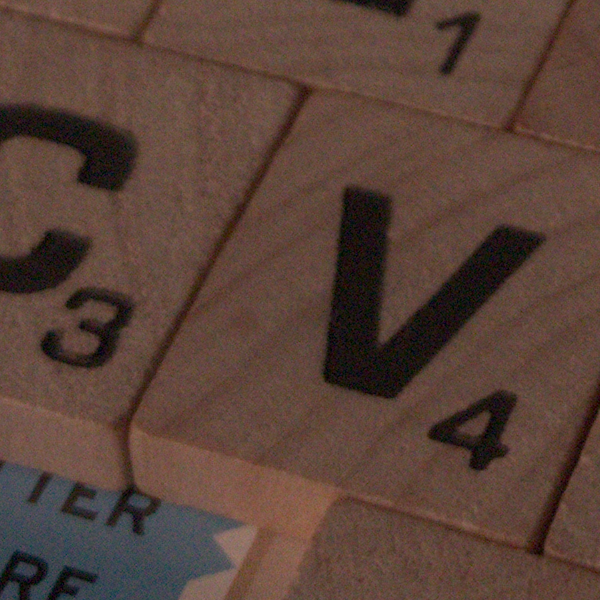}{31.64dB}  &
	\overimgx[width=0.23\linewidth]{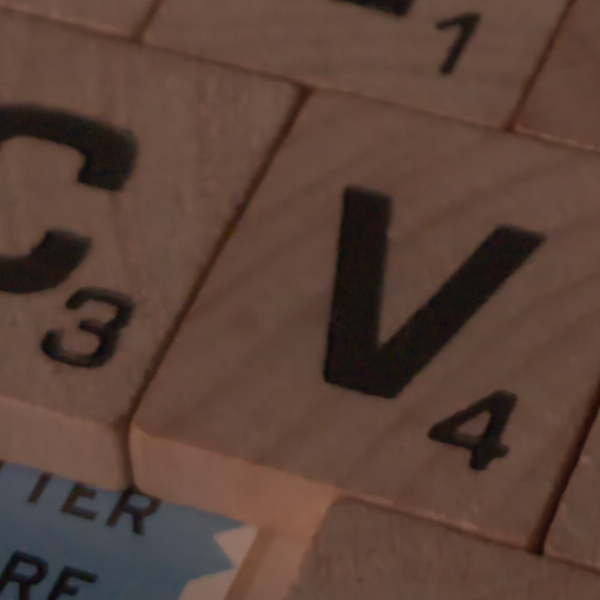}{41.37dB}  &
	\overimgx[width=0.23\linewidth]{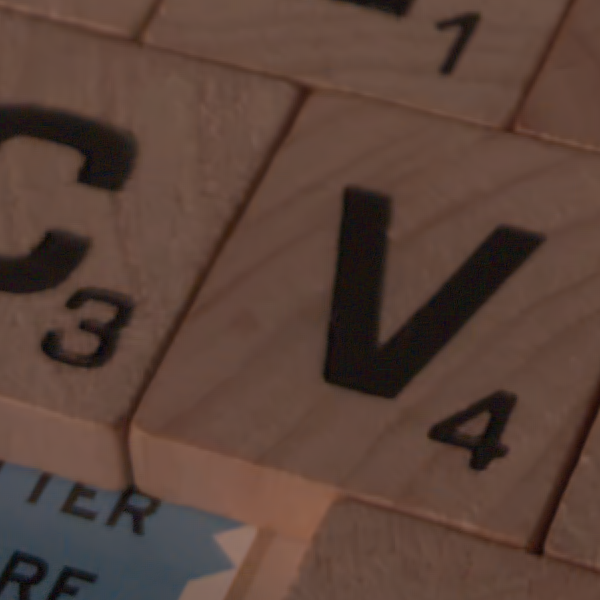}{41.34dB}  \\
	noisy demosaiced & JCNN~\cite{gharbi2016deep} & cfaBM3D+MLRI \\
	& & ($\A\&\B$) \\
	
	\overimgx[width=0.23\linewidth]{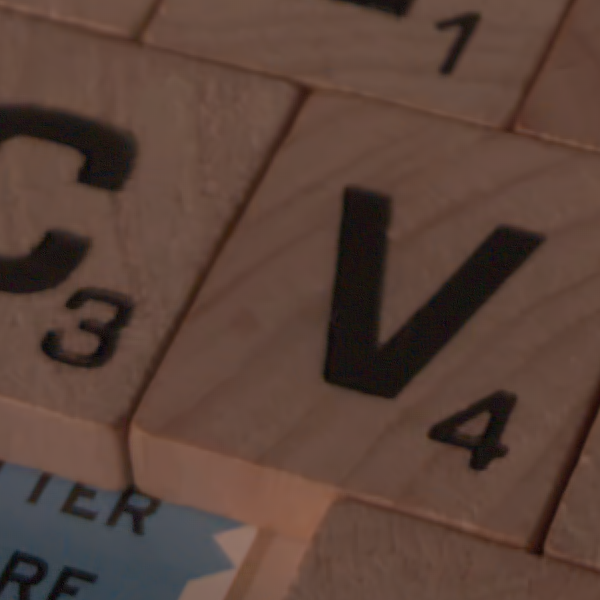}{41.61dB}  &
	\overimgx[width=0.23\linewidth]{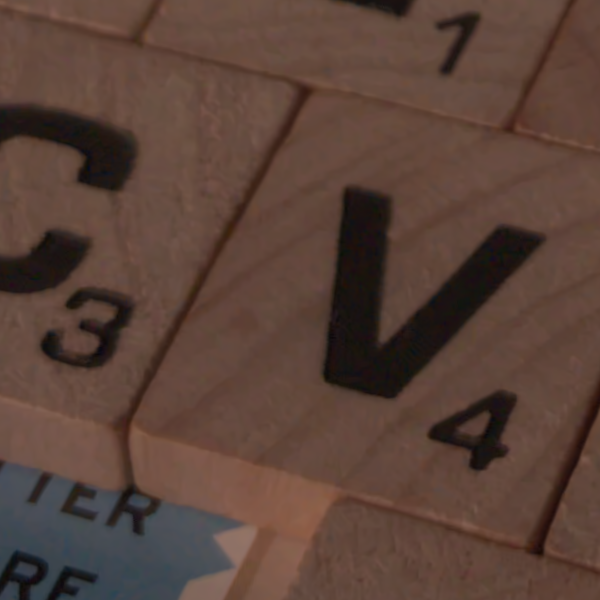}{41.66dB}  &
	\overimgx[width=0.23\linewidth]{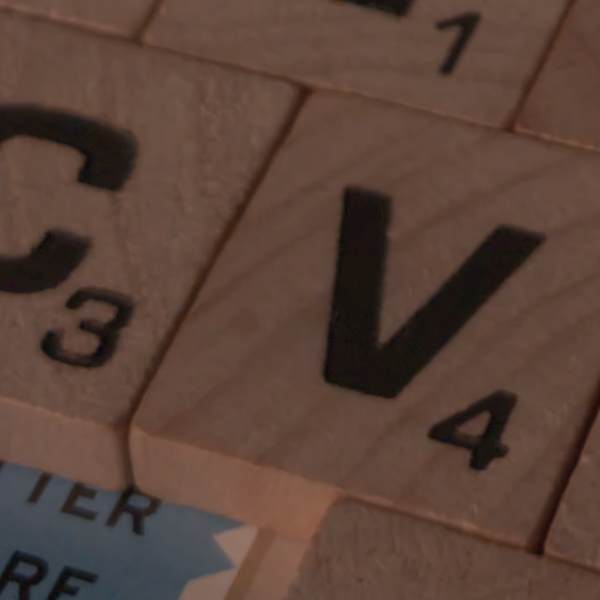}{42.80dB}
	\\   
	 cfaBM3D+RCNN & MLRI+CBM3D &  RCNN+CBM3D \\
	 ($\A\&\B$) & ($\B\&1.5\A$) & ($\B\&1.5\A$) \\
	\end{tabular}
	\caption{Demosaicing and denoising results on an image from the SIDD dataset. 
	We compare the two schemes of $\A\&\B$, cfaBM3D+MLRI and cfaBM3D+RCNN, the two schemes of $\B\&1.5\A$, MLRI+CBM3D and RCNN+CBM3D. As a reference we also include the result of JCNN, a joint CNN method.}
	\label{fig SIDD}
\end{figure*}

Table~\ref{tab:real images} shows the CPSNR and estimated noise levels of images generated by different schemes on the SIDD dataset. We list them separately by phone model.  It can be seen from   Table~\ref{tab:real images} that the $\B\&1.5\A$ solution is more competitive than the $\A\&\B$ solution in terms of CPSNR, with an average 0.60 dB gain. This is consistent with the previous results on the simulated data. Figure~\ref{fig SIDD} shows the visual quality of both strategies. 
 JCNN is not competitive on the SIDD dataset, because it  was not trained on this dataset. This also shows that our proposed scheme has better robustness and adaptability than  JCNN.
The $\B\&1.5\A$ scheme  keeps more image details than others.

In a word, the $\B\&1.5\A$ scheme clearly outperforms  $\A\&\B$  in visual quality  and numerical results for both simulated data and real data. Our results also provide theoretical support for real sRGB image denoising which removes noise from full color images after demosaicing. The next section addresses raw image denoising.

\begin{figure}
	\begin{center}
			\includegraphics[width=0.8\linewidth]{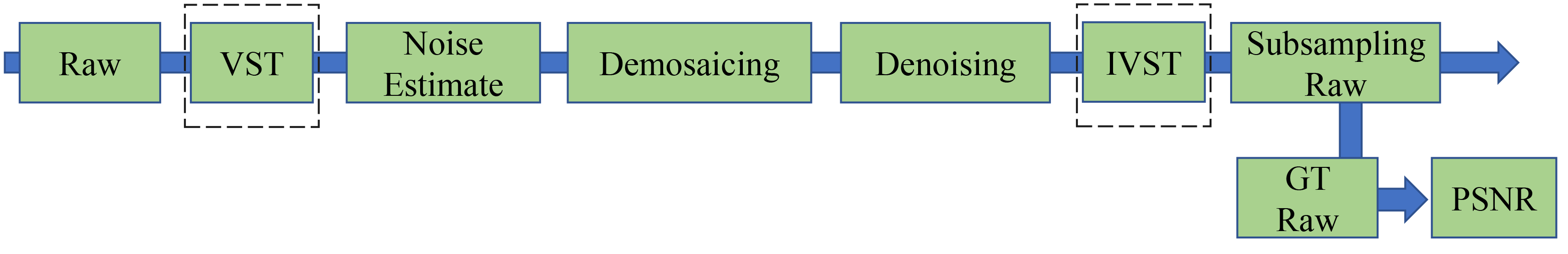}
		\caption{The flowchart of raw image denoising under $\B\&1.5\A$ scheme. The dashed VST/IVST blocks are active in just one of the pipeline variants.}
		\label{Raw_Dn}
	\end{center}
\end{figure}

\begin{table}[t]
	\caption{Validation of the $\B\&1.5\A$ scheme on the SIDD dataset. Note that for each camera, images with different noise levels are being considered. The noise range is $\sigma \in [0.48, 22.59]$ without VST and $\sigma\in [0.38, 13.00]$ with VST. The best result is in \textcolor{red}{red}, the second best one is in \textcolor{brown}{brown}.
	}
	\label{SIDD_DN}
    \begin{center}
	\renewcommand{\arraystretch}{1.1} \addtolength{\tabcolsep}{-1pt}
	\footnotesize
    \begin{tabular}{cc|cccccc}
                             &        & cfaBM3D & JCNN & HA+    & RCNN+  & RCNN+ & MLRI+\\
                             &        &       &      & CBM3D  & FFDNet & CBM3D & CBM3D\\ \hline 
        \multirow{2}{*}{Raw} & VST    & 49.03 & 46.05 & 49.18 & 48.51 & \textcolor{brown}{49.30} & \textcolor{red}{50.55} \\ 
                             & non-VST & 48.53 & 45.51 & 49.02 & 48.55 & \textcolor{brown}{49.22} & \textcolor{red}{50.45}  \\ 
                             \\
    \end{tabular}
    \end{center}
\end{table}
\begin{table}[t]
	\caption{Comparison results of the $\B\&1.5\A$ scheme on the SIDD and DND benchmarks (results as reported on the corresponding websites).
	* indicates the use of the variance stabilizing transform (VST).  
	The best result is in \textcolor{red}{red}, and the second best one is in \textcolor{brown}{brown}. 
	}
	\label{test Benchmark}
    \begin{center}
	\renewcommand{\arraystretch}{1.1} \addtolength{\tabcolsep}{-1pt}
	\footnotesize
    \begin{tabular}{l|ccccccc|c}
    \multirow{2}{*}{Raw} & TNRD  & MLP   & EPLL  & WNNM  & BM3D  & RCNN+  & MLRI+ & CycleISP \\ 
                          &       &       &       &       &       & CBM3D & CBM3D & \\\hline
     SIDD & 42.77 & 43.17 & 40.73 & 44.85 & 45.52 & \textcolor{brown}{48.36}& \textcolor{red}{49.43}  & 47.98 \\
     SIDD* & -- & -- & -- & -- & -- & \textcolor{brown}{48.56} & \textcolor{red}{49.48} &  --\\
     DND  & 44.97 & 42.70 & 46.31 & 46.30 & 46.64 & \textcolor{brown}{47.16} & \textcolor{red}{47.63} & \textbf{49.13} \\
     DND* & 45.70 & 45.71 & 46.86 & 47.05 & 47.15 & \textcolor{brown}{47.26} & \textcolor{red}{47.76} & --\\
    \end{tabular}
    \end{center}
\end{table}

\subsection{The \texorpdfstring{$\B\&1.5\A$}{} strategy for raw image denoising}
\label{Sec: raw image denoising}

We applied the $\B\&1.5\A$ scheme to raw image denoising. To that aim, we defined the pipeline shown in Figure~\ref{Raw_Dn}. We considered two pipeline variants: with and without variance stabilizing transform.
In the first case, a variance stabilizing transformation was used to transform the raw image noise into approximate Gaussian noise, and the noise level in each image was then estimated by the method~\cite{Chen2015_noise_level}. In the second case, we applied the noise estimation method~\cite{Chen2015_noise_level} directly on the original noise images. 
Table~\ref{SIDD_DN} shows the results of the $\B\&1.5\A$ scheme on the raw images of the SIDD dataset~\cite{Abdelhamed2018}. Note that applying the VST leads to slightly better results in almost all cases.
RCNN underperforms when handling raw data, because its training data is sRGB data. MLRI is a traditional interpolation algorithm, which is not affected by different color spaces and achieves the best results.
The estimated noise range for the original noisy images in the \textbf{SIDD raw image datasets} is $\sigma\in [0.48, 22.59]$ and after VST is $\sigma\in [0.38, 13.00]$. 
According to Table~\ref{table_cma}, the results of the CMA-ES optimized scheme and the $\B\&1.5\A$ scheme are almost equal when the noise level $\sigma \leq 20$, which justifies the use of $\B\&1.5\A$ (more precisely, the noise level of all considered images is always less than $23$).
Considering the trade-off between reconstruction quality and computational consumption, the $\B\&1.5\A$ scheme is more valuable for the considered application.

\begin{figure*}[t]
	\footnotesize
	\centering
	\addtolength{\tabcolsep}{-4pt} 
	
	\begin{tabular}{ccc}
		\overimgxx[width=.28\linewidth,trim={0 0 0 0},clip]{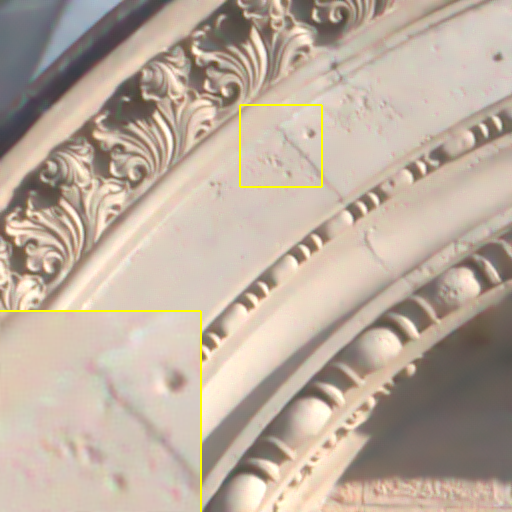}{36.90dB}
		& 
		\overimgxx[width=.28\linewidth,trim={0 0 0 0},clip]{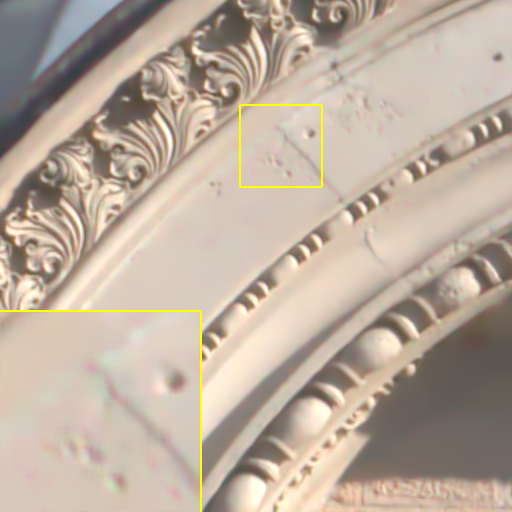}{38.20dB} 
		&
		\overimgxx[width=.28\linewidth,trim={0 0 0 0},clip]{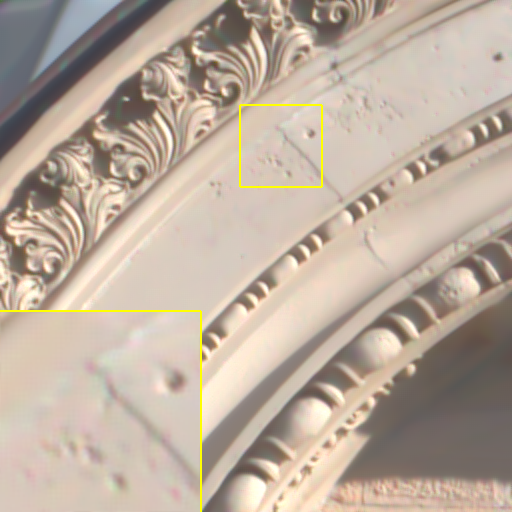}{38.11dB} \\
		TNRD~\cite{Chen2015TNRD} & EPLL~\cite{Zoran2011EPLL} & WNNM~\cite{gu2014weighted} \\
		
		\overimgxx[width=.28\linewidth,trim={0 0 0 0},clip]{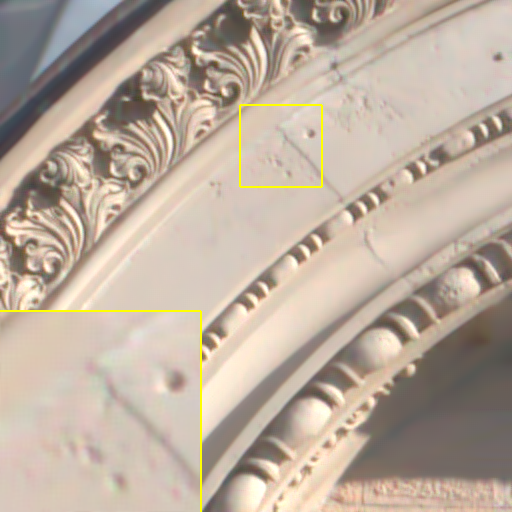}{37.84dB}
		&
		\overimgxx[width=.28\linewidth,trim={0 0 0 0},clip]{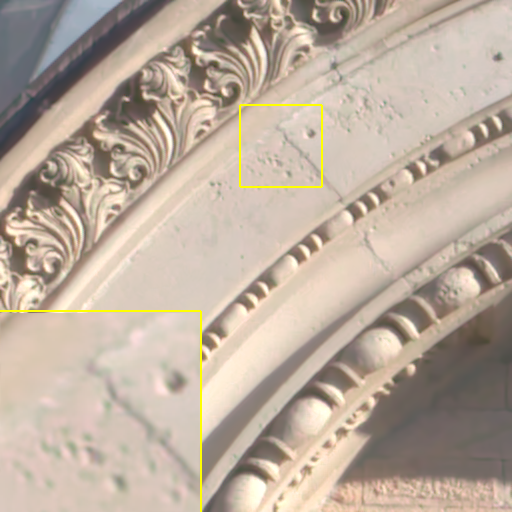}{38.44dB}
		&
		\overimgxx[width=.28\linewidth,trim={0 0 0 0},clip]{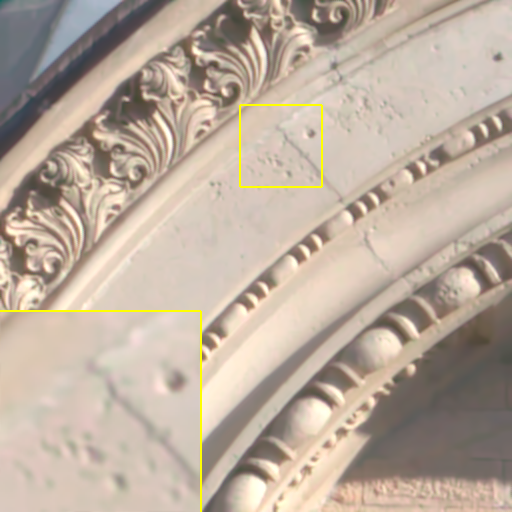}{40.07dB}\\ 
		 BM3D~\cite{Dabov2007BM3D} & RCNN+1.5CBM3D & MLRI+1.5CBM3D \\
		
		\overimgxx[width=.28\linewidth,trim={0 0 0 0},clip]{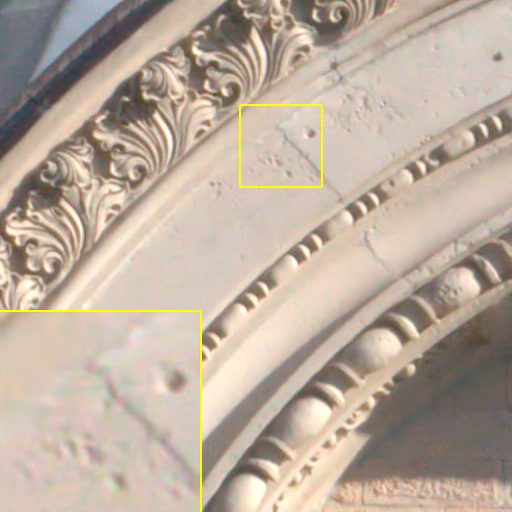}{36.91dB}
		& 
		\overimgxx[width=.28\linewidth,trim={0 0 0 0},clip]{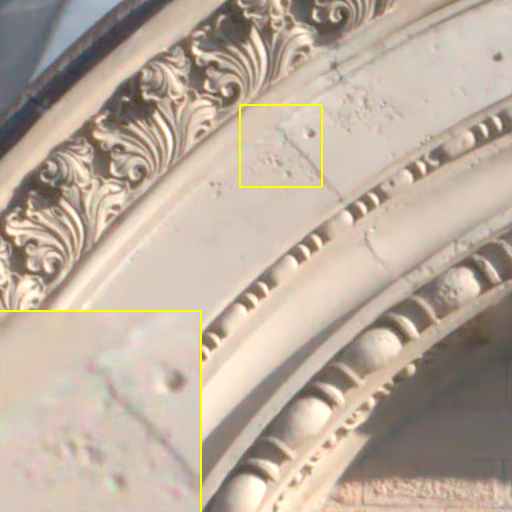}{36.77dB} 
		&
		\overimgxx[width=.28\linewidth,trim={0 0 0 0},clip]{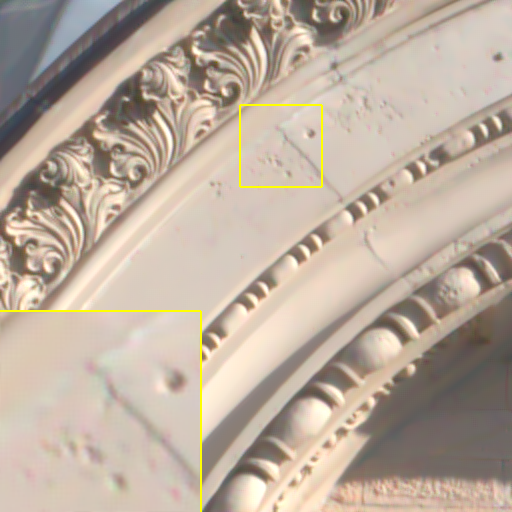}{38.00dB} \\
		TNRD~\cite{Chen2015TNRD} (VST) & EPLL~\cite{Zoran2011EPLL} (VST) & WNNM~\cite{gu2014weighted} (VST) \\
		
		\overimgxx[width=.28\linewidth,trim={0 0 0 0},clip]{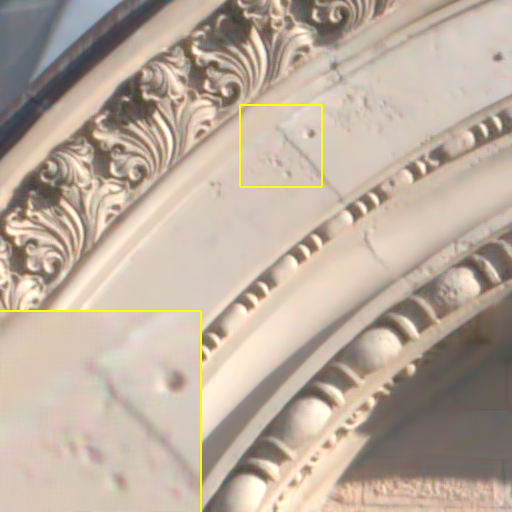}{37.53dB}
		&
		\overimgxx[width=.28\linewidth,trim={0 0 0 0},clip]{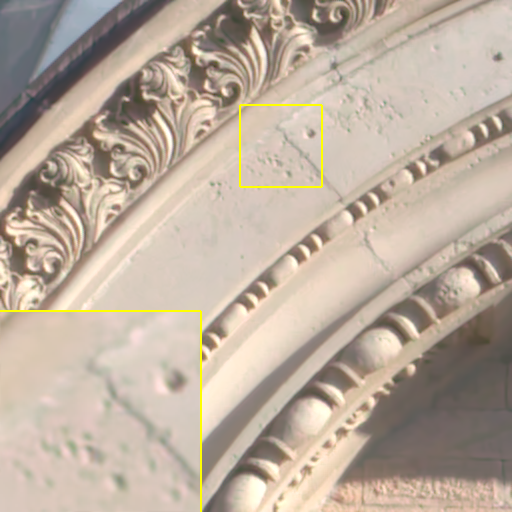}{38.53dB}
		&
		\overimgxx[width=.28\linewidth,trim={0 0 0 0},clip]{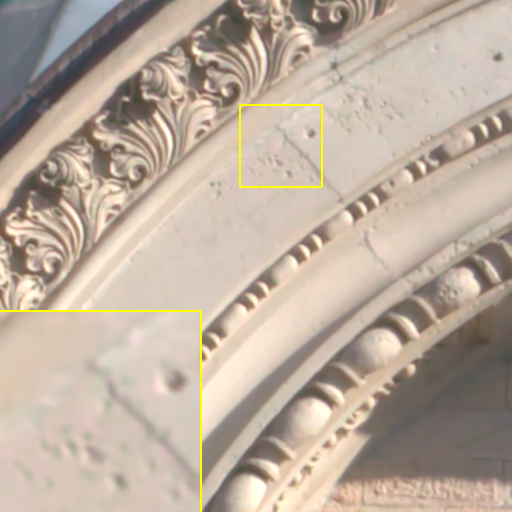}{40.16dB}\\ 
		
		 BM3D~\cite{Dabov2007BM3D} (VST) & RCNN+1.5CBM3D (VST) & MLRI+1.5CBM3D (VST)\\
		
	\end{tabular}
	\caption{Denoising results on an image from the DND dataset. We compare the $\B\&1.5\A$ scheme  (MLRI+CBM3D and RCNN+CBM3D), TNRD~\cite{Chen2015TNRD}, EPLL~\cite{Zoran2011EPLL}, WNNM~\cite{gu2014weighted} and BM3D~\cite{Dabov2007BM3D} (results as reported on the benchmark website). 
		}
	\label{fig:DND}
\end{figure*}

To further validate the performance of the $\B\&1.5\A$ scheme, we compared MLRI+CBM3D and RCNN+CBM3D with TNRD~\cite{Chen2015TNRD}, EPLL~\cite{Zoran2011EPLL}, WNNM~\cite{gu2014weighted}, BM3D~\cite{Dabov2007BM3D} and CycleISP~\cite{Zamir2020CycleISP} on the SIDD~\cite{Abdelhamed2018} and DND~\cite{2017DND} benchmarks.
As with the previous results, the noise ranges of the \textbf{raw images} in the SIDD and DND benchmarks are respectively $\sigma\in [0.57, 21.39]$ and $\sigma\in [0.59, 14.97]$, and after VST the noise ranges are $\sigma\in [0.46, 12.79]$ and $\sigma\in [0.44, 9.17]$, which still satisfy the best use case for $\B\&1.5\A$.
The relevant results are shown in Table \ref{test Benchmark}, and more detailed results can be found on the SIDD\footnote{ \url{http://www.cs.yorku.ca/~kamel/sidd/benchmark.php} } and DND\footnote{ \url{https://noise.visinf.tu-darmstadt.de/benchmark/\#results_raw} } websites.
The CycleISP result is better on DND than our best proposed scheme MLRI+CBM3D, but not on SIDD, this is likely due to the domain difference between DND and SIDD (as SIDD has darker images).  Therefore, this deep learning based approach has several caveats: first MLRI and CBM3D offer guarantees of domain independence and were not trained on the specific image pipeline associated with DND. Second, a difference of 1.5 dB is anyway visually invisible for such high PSNRs as those involved in the table (see Figure~\ref{fig:DND}). 
Third, MLRI and CBM3D can be accelerated without  performance loss on dedicated architectures while the computational weight of a CNN is hardly reducible.

Although the $\B\&1.5\A$ scheme falls short of state-of-the-art deep learning raw image denoising methods such as CycleISP~\cite{Zamir2020CycleISP}, our proposed lightweight scheme is still the best among traditional algorithms and it even outperforms some deep learning algorithms (see the DND benchmark website).
Compared to the computational resources consumed by deep learning methods, our proposed scheme is computationally very competitive. 
Figure~\ref{fig:DND} shows the comparison of the visual quality of traditional algorithms on raw image denoising. 
Our scheme keeps more details, introduces fewer color artifacts than other traditional algorithms and avoids checkerboard artifacts.
With a lightweight demosaicker, BM3D obviously improves on raw image denoising with an average gain of 3.91 dB for SIDD, 0.99 dB for DND and 0.61 dB for DND with VST. As a result, we can conclude that the $\B\&1.5\A$ scheme is very effective for raw image denoising.

\subsection{Time consumption and generalizability}

\begin{table}[t]
	\caption{Time consumption. The average running time (CPU) of the three strategies in processing 10 images on a PC with an Intel Core i7-9750H 2.60GHz CPU and 16GB memory. Note that we do not use the deep learning methods and only compared the traditional methods.
	}
	\label{time}
    \begin{center}
	\renewcommand{\arraystretch}{1.1} \addtolength{\tabcolsep}{-1pt}
	\footnotesize
    \begin{tabular}{|c|c|c|c|c|c|c|} 
    \hline 
    \multicolumn{3}{|c|}{$\A\&\B$} & \multicolumn{3}{c|}{$\B\&1.5\A$} & CMA-ES  \\ \hline
     cfaBM3D+ & cfaBM3D+ & cfaBM3D+ & HA+   & RI+   & MLRI+ & cfaBM3D+ \\ \ 
     HA~      & RI~      & MLRI     & CBM3D & CBM3D & CBM3D & MLRI+    \\
              &          &          &       &       &       & CBM3D    \\ \hline 

    7.41 s    & 7.64 s   & 7.85 s   & 16.16 s & 16.66 s & 16.72 s & 23.93 s   \\ \hline 
    \end{tabular}
    \end{center}
\end{table}

We examined the runtimes of three strategies and evaluated the generalizability of the CMA-ES scheme, aiming to achieve a balance between good performance and reasonable runtimes. We limited our comparison to traditional algorithms, as deep learning algorithms require long computing times on CPUs. 
Table \ref{time} shows the running times of the three strategies on a PC with an Intel Core i7-9750H 2.60GHz CPU and 16GB memory.  
As the table demonstrates, the demosaicing algorithm has a negligible runtime, while the majority of the computational time is spent on denoising. 
The computation time of $\A\&\B$ is half that of $\B\&1.5\A$, because $\A\&\B$ processes two half-size images, which is exactly half the size of the full-color images processed by $\B\&1.5\A$.  
In terms of the trade-off between time consumption and performance, $\B\&1.5\A$ is the optimal choice, particularly for moderate levels of noise ($\sigma \leq 20$, as described in Section \ref{Sec: raw image denoising}).
However, for high noise scenes, the $\A_{1}\&\B\&\A_{2}$ pipeline may be the best option for achieving optimal performance.

\begin{table}[t]
	\caption{Generalizability of CMA-ES optimal parameters to different noise levels. Evaluation of noise levels with $\sigma = 50$ proximity (selected as 46 to 54) using two generalization schemes.
	}
	\label{generalizability}
    \begin{center}
	\renewcommand{\arraystretch}{1.1} \addtolength{\tabcolsep}{0pt}
	\footnotesize
    \begin{tabular}{c|cccc} 
     $\sigma$ & $\A\&\B$ & $\B\&1.5\A$ & CMA-ES & CMA-ES  \\ 
     & & & image transformation & $\sigma$ transformation   \\ \hline
     46 & 24.10 & 24.60 & 24.83 & \textbf{24.90} \\ 
     47 & 23.98 & 24.46 & 24.74 & \textbf{24.78} \\ 
     48 & 23.85 & 24.32 & 24.63 & \textbf{24.64} \\ 
     49 & 23.74 & 24.19 & \textbf{24.52} & \textbf{24.52} \\ 
     51 & 23.50 & 23.91 & \textbf{24.26} & \textbf{24.26} \\ 
     52 & 23.35 & 23.77 & \textbf{24.13} & 24.12 \\
     53 & 23.24 & 23.64 & \textbf{24.00} & \textbf{24.00} \\
     54 & 23.14 & 23.52 & \textbf{23.90} & 23.89 \\ 
    \end{tabular}
    \end{center}
\end{table}

We now turn our attention to the generalization of the CMA-ES optimization parameters, which requires a large number of calculations, making the optimization process time-consuming. 
One critical aspect is the independence of the parameters from the dataset. 
This issue arises implicitly in the previous discussion. In Section \ref{sec:pipeline}, we employed the Imax dataset for the CMA-ES optimization, whereas the parameters were applied directly to the Kodak dataset in the comparison (see Tables \ref{table_cma} and \ref{Table KodakCompare}).
As demonstrated in these tables, the CMA-ES optimal parameters remain consistent when applied to the Kodak dataset, which leads to the conclusion that the CMA-ES optimization parameters exhibit good generalization across datasets.

Another crucial aspect is the generalization to different noise levels. Given that it is impractical to train optimal parameters each time for real-world applications, it is essential to discuss what to do when the noise level does not match the level of optimal parameters.  We propose two schemes: 
\begin{itemize}
    \item  Image transformation, where the image is transformed to the nearest noise level using the corresponding optimal parameters $\alpha,\beta,\sigma_{1},\sigma_{2}$, namely $\frac{x}{\sigma^{*}}\sigma$ and its inverse $\frac{y}{\sigma}\sigma^{*}$, where $x$ is the noisy image, $y$ is the reconstructed image, $\sigma^{*}$ is the actual noise level, and $\sigma$ is the nearest noise level with known optimal parameters;

    \item $\sigma$ transformation, where the optimal parameter $\alpha,\beta$ for the nearest noise level is directly used, and the parameters $\sigma_{1}$ and $\sigma_{2}$ are transformed by $\sigma_{1}^{*} = \frac{\sigma_{1}}{\sigma}\sigma^{*}$ and $\sigma_{2}^{*} = \frac{\sigma_{2}}{\sigma}\sigma^{*}$, where $\sigma^{*}$ is the actual noise level, and $\sigma$ is the nearest noise level with known optimal parameters.
    
\end{itemize}
We evaluated the how both schemes generalize around $\sigma = 50$ (selected as 46 to 54). The corresponding results are presented in Table \ref{generalizability}. As shown in the table, both  schemes outperform the $\A\&\B$ and $\B\&1.5\A$ strategies, indicating the generality of the CMA-ES optimization parameters over a range without the need for repeated optimization.

From Table \ref{time}, it is apparent that the denoising stage is responsible for the majority of the time consumption. Therefore, it is advisable to use a fast algorithm, such as the BM3D algorithm implemented on the GPU \cite{Davy2021BM3D}  when using the CMA-ES algorithm to obtain optimal parameters.

\section{Conclusion}
\label{sec:conclusion}

This paper  established a model to optimize the denoising and demosaicing pipeline. 
The optimal pipeline (obtained by CMA-ES) is a $\A_{1}\&\B$ $\&\A_{2}$ scheme with appropriate parameters and $\B\&1.5\A$ is almost equal to the optimal one when $\sigma \leq 20$.
Our best performing  combination in terms of  quality  and speed is a $\B\&1.5\A$ scheme for two reasons: the  $\A_{1}\&\B\&\A_{2}$ scheme gets the best result, but it takes twice as many calculations as $\B\&1.5\A$; as discussed in Section \ref{Sec: raw image denoising}, in most cases, the noise level for raw images is less than 20.  
Experiments show a considerable gain.  
The results of the $\B\&1.5\A$ scheme show a  0.5 to 1 dB gain, when compared with the best $\A\&\B$ strategy.  These conclusions apply for moderate noise ($\sigma \leq 20$) but remain valid for high noise, where we nevertheless found a slight improvement of about 0.3 dB for a twice more complex pipeline  $\A_{1}\&\B\&\A_{2}$  with two denoising steps.
We also gave a detailed theoretical explanation of why the $\B\&1.5\A$ scheme is superior to the $\A\&\B$ scheme.

We also saw that, unsurprisingly, heavy weight learning-based joint demosaicing and denoising achieves the best performance. However, the above conclusions are still crucial for practical light weight and domain independent application scenarios. They might also inspire the design and training of deep learning algorithms.

\section*{Acknowledgment}
This work was supported by National Natural Science Foundation of China (No.~12061052), Natural Science Fund of Inner Mongolia Autonomous Region (No.~2020MS01002), Young Talents of Science and Technology in Universities of Inner Mongolia Autonomous Region (No. NJYT22090), Innovative Research Team in Universities of Inner Mongolia Autonomous Region (No. NMGIRT2207), Prof. Guoqing Chen's “111 project” of higher education talent training in Inner Mongolia Autonomous Region, Inner Mongolia University Postgraduate Research and Innovation Programmes (No. 11200-5223737), the network information center of Inner Mongolia University, Office of Naval research grant N00014-17-1-2552, DGA Astrid project n$^\circ$ ANR-17-ASTR-0013-01. 
Y. Guo and Q. Jin are very grateful to Professor Guoqing Chen for helpful comments and suggestions. 
The authors are also grateful to the reviewers for their valuable comments and remarks.

\bibliographystyle{AIMS}
\bibliography{ref}

\medskip
Received xxxx 2022; revised xxxx 2023; early access xxxx 20xx.
\medskip

\end{document}